\documentclass[epj]{svjour}
\usepackage{graphicx}
\usepackage{dcolumn}
\usepackage{epsfig,amsmath}
\graphicspath{{pics/}}    

\def\NP{N_\mathrm{Part}}
\def\Nbin{N_\mathrm{bin}}

\def\bi{\begin{itemize}}
\def\ei{\end{itemize}}
\def\D{$D$}

\def\KKbar{$K^+K^-$}
\def\KKbarnul{$K^0\bar{K^0}$}

\def\d{{\rm d}}
\def\e{{\rm e}}

\def\fm3{$\mathrm{fm}^3$}
\def\f3{\mathrm{fm}^3}

\def\be{\begin{equation}}
\def\ee{\end{equation}}

\def\jpsi{J/$\psi$}
\def\psipr{$\psi$'}
\def\sqrtsnn{$\sqrt{s}_\mathrm{NN}$}

\begin{document}

\title{Dilepton production in $p+p$, $Cu+Cu$ and $Au+Au$ collisions
at $\sqrt{s}$ = 200 GeV}
\author{J. Manninen\inst{1}, E. L. Bratkovskaya\inst{2},
W. Cassing\inst{1}, O. Linnyk\inst{2}}
\institute{Institut f\"{u}r Theoretische Physik,
      Universit\"{a}t Giessen, 35392 Giessen, Germany
\and  Institut f\"{u}r Theoretische Physik,
      Universit\"{a}t Frankfurt, 60054 Frankfurt, Germany}

\abstract{
We study dilepton production in proton-proton, $Cu+Cu$ as well as
in $Au+Au$ collisions at the center-of-mass energy $\sqrt{s}$= 200
GeV per participating nucleon pair within an extended statistical
hadronization model. In extension to earlier studies we incorporate
transport calculations for an estimate of uncorrelated
$\e^+\e^-$ -pairs from semileptonic \D~meson decays. While the invariant mass
spectrum of dielectrons is well understood in the $p+p$
collisions, severe discrepancies among different model scenarios
based on hadronic degrees of freedom and recent data from the
PHENIX Collaboration are found in heavy-ion collisions in the low mass 
region from 0.15 to 0.6 GeV as well as in the intermediate mass regime 
from 1.1 to 3 GeV when employing the standard dilepton sources. We 
investigate, furthermore, the background from correlated dileptons that 
are not emitted as a pair from a parent hadron but emerge from semileptonic 
decays of two correlated daughter hadrons. Our calculations suggest a 
sizeable contribution of such sources in central heavy-ion collisions in 
the low mass region. However, even the upper limits of our calculations are 
found to be far below the dilepton mass spectra of the PHENIX Collaboration.
\\}


\PACS{
   {25.75.-q}{} \and
   {13.60.Le}{} \and
   {14.40.Lb}{} \and
   {14.65.Dw}{}
}

\authorrunning{Manninen \emph{et al.}}

\maketitle

\date{Revised version: 18th March 2011}
%

\section{Introduction}

The goal of this work is to study the production of correlated
electron-positron pairs in proton-proton and heavy-ion collisions
at \sqrtsnn=200 GeV. We will evaluate the invariant mass spectrum
${\d N}/{\d m_{e^+e^-}}$ within an extended hadronization model
up to invariant masses of 4 GeV thus covering the
low mass as well as the intermediate mass and charmonium regime.
We confront our calculations with corresponding measurements of
the PHENIX collaboration (within the experimental acceptance)
which has taken data \cite{:2008asa,Adare:2009qk} up to 4 GeV in
$m_{e^+e^-}$ at mid-rapidity.  In the case of proton-proton
collisions the PHENIX collaboration has found out that the
measured spectrum can be described very well up to masses of 4
GeV, if one takes into account all relevant hadronic sources of dileptons
in the analysis. This was done by a simultaneous measurement of all hadron
rapidity densities and transverse momentum spectra around
mid-rapidity;  by using these experimental rapidity densities one
can then estimate the dilepton yields at different invariant
masses due to the known hadronic decays to $e^+e^-$ pairs.
Independently, microscopic transport calculations within the
Hadron-String-Dynamics (HSD) framework~\cite{Ehehalt:1996uq,Cassing:1999es}
have come to the same conclusion when incorporating the measured cross
section for
$c\bar{c}$ pairs from the PHENIX collaboration \cite{lena2009}.

In contrast to the proton-proton collisions, the measured
invariant mass spectrum of dileptons in $Au+Au$ collisions have so
far not been properly understood theoretically. Instead,
theoretical estimates are found to deviate up to a factor of 4 or
5 from the PHENIX data for central Au+Au collisions in the low mass regime
\cite{lena2009,Dusling} (cf. also Fig. 42 in Ref. \cite{Adare:2009qk})
and by up to a factor 2 to 3 for intermediate masses. Such
discrepancies among (hadronic) models and experimental data have
not been observed at lower Super-Proton-Synchrotron (SPS) beam
energies and in different collision systems
\cite{lena2009,vanHees:2007th} where a major broadening of the
vector-meson resonances is reported. This might suggest that
non-hadronic dilepton channels could be responsible for the
discrepancies observed so far. This issue needs further
independent investigations.

In this work, we will partly repeat the analysis by the PHENIX collaboration
in the intermediate mass
range especially with respect to the contribution from charmed meson decays
but instead
of using the measured yields as input for the dilepton sources, we wish to
implement different models in order to calculate the rapidity densities of
different hadrons and subsequently estimate the emission of dileptons from
their decays. These models are  controlled by
the PHENIX data for $p+p$ collisions.

We recall that the
invariant mass spectrum of dileptons can be divided in three fairly
distinct regions in each of which different physics processes are dominant.
Below the $\phi$ meson mass, the region referred hereafter as the low mass
region (LMR: $m_{e^+e^-}\in$ [0.0 ; 1.1] GeV), the
dilepton production is dominated by the decays of non-charmed mesons,
i.e. mesons with essentially light quark content ($u,d,s$). In
the intermediate mass region (IMR: $m_{e^+e^-}\in$[1.1 ; 3.2] GeV),
i.e. in between the $\phi$ meson and \jpsi~mass, the invariant mass spectrum of
electron-positron pairs is dominated by the semileptonic decay products of open
charm mesons. Strictly speaking this is just a background for the ''true''
dilepton sources, but since this component is always present in the analysis,
one needs to carefully evaluate the contribution from open charm as well.
Furthermore, above about 3 GeV of invariant mass the
direct decays of charmonia become dominant and provide a
constraint on the number of produced $c\bar{c}$-pairs - forming bound states
- once the charmonium suppression is controlled independently.
We concentrate in this article on studying the LMR and IMR  thus exceeding
previous
approaches that focused on the LMR and extrapolated to the IMR. The high mass
region (HMR:$m_{e^+e^-}>$3.8 GeV) of the dilepton invariant mass spectrum is
dominated by the Drell-Yan process and $B$ meson decays which we will not
address here.

In describing the yields or
ratios of particle yields of hadrons consisting of $u, d$ and $s$ quarks
phenomenological models, in particular ``thermal models'', have
proven to be very useful due to their simplicity and low number of adjustable
parameters. In this work, we will evaluate the yields of light mesons
within the statistical hadronization model which has been applied to
high-energy elementary~\cite{Kraus:2007hf,Becattini:2008tx,Andronic:2009sv}
and especially
heavy-ion~\cite{Becattini:2000jw,Cleymans:2001at,Baran:2003nm,Cleymans:2004pp,Rafelski:2004dp,Andronic:2005yp,Becattini:2005xt,Letessier:2005qe,Manninen:2008mg,Andronic:2008gu,NoronhaHostler:2010yc}
collision experiments in order to calculate
the yields of different hadron species with fairly a lot of
success. We mention that the CERES Collaboration has also analyzed
their dilepton data in the LMR on the basis of the statistical hadronization
model with some success at Super-Proton-Synchrotron (SPS)
energies for $Pb+Au$ collisions~\cite{Agakichiev:2005ai}.

Unlike for the bulk meson production in the LMR, the statistical
hadronization model can not be applied to estimate the yields of charmed
hadrons in the IMR. Instead,
we need to rely on experimental information here in order to estimate
the differential yields of charmed hadrons. To this aim we will
formulate an 'extended statistical model' including early
charm-pair production, collective flow of the hot and dense matter
as well as correlated and uncorrelated semileptonic decays of
\D~mesons. The amount of \D~meson rescattering will be followed
up by the HSD transport approach~\cite{Linnyk:2008hp} in order to estimate
the amount of surviving correlated semileptonic decays in the heavy-ion
collisions.

In extension to earlier studies we will, furthermore, explore the
contribution of correlated semileptonic decays stemming from
kaon-antikaon pairs that emerge from the decay of heavy parent
hadrons. The most important of these sources are proportional to $\gamma_S^2$,
where $\gamma_S$ denotes the strangeness suppression factor which is low in
proton-proton reactions but close to unity in nucleus-nucleus
collisions. Our upper limits for these contribution will be
compared to the PHENIX data within the proper acceptance windows
as well as the conventional dilepton sources mentioned above.

\section{An extended statistical hadronization model}\label{SHM}

The statistical hadronization model (SHM) has been applied
successfully in calculating the number of emitted hadrons in
high-energy collision systems~\cite{Kraus:2007hf,Becattini:2008tx,Andronic:2009sv,Becattini:2000jw,Cleymans:2001at,Baran:2003nm,Cleymans:2004pp,Rafelski:2004dp,Andronic:2005yp,Becattini:2005xt,Letessier:2005qe,Manninen:2008mg,Andronic:2008gu,NoronhaHostler:2010yc}.
This model is
well documented in the references given above and accordingly we
will introduce the main concepts only. We evaluate the hadron
yields in the grand-canonical ensemble because the calculations
simplify substantially if one does not require exact conservation of
Abelian charges and/or energy-momentum. Our choice is motivated by
two reasons: First of all experimental observations show that the
grand-canonical ensemble is sufficient enough, i.e. that at RHIC energies
the data are well described under the approximations we have chosen. The
other reason is that in order to evaluate the hadron yields in the
canonical ensembles, one needs to know the volume as well as the
exact (integer) charges on an event by event basis. However, the
PHENIX collaboration has measured only a small fraction of the emitted hadrons
while a large part of the system is never observed. We would need
to make severe assumptions for the part of the system not
measured, if we were to implement the canonical formalism for the
calculation. We also note that the canonical effects are most
pronounced for heavy and exotic hadrons, while the bulk of the
dilepton emission arises from the low-mass mesons, which are
produced abundantly and do not suffer from canonical suppression
effects. Thus, we deem that performing the analysis in the
grand-canonical ensemble should be good enough for our purposes
and indeed this seems to be confirmed by our work (see below).

In the SHM, the primary hadron multiplicity of hadron type $i$
is calculated (in the on-shell Boltzmann approximation) according
to
\begin{equation} \label{dndy}
N_i = V
\frac{2J_i+1}{(2\pi)^3}
\int \gamma_S^{n_s} \ \e^{{\bf \mu}\cdot {\bf q}_i/T} \
\e^{-\sqrt{p^2 + m_i^2}/T} \ \d^3p.
\end{equation}
In Eq. (\ref{dndy}) $J_i$ denotes the spin, $p$ the momentum and $m_i$
the mass of
the particle while $q_i$ is a vector consisting of the baryon,
electric and strangeness charges of the hadron species $i$.
The state of the ''thermal'' fireball is specified by its
temperature $T$, volume $V$ and chemical potentials
(collected in the vector $\mu$) for baryon, electric and strangeness
charges.

Several independent experimental measurements have verified that
the mid-rapidity region is actually almost net charge free in
proton-proton and $Au+Au$ collisions at \sqrtsnn=200
GeV~\cite{Back:2002ks,Adams:2003xp,Bearden:2003fw}.
The baryon chemical potential is expected to be of the order of
30 MeV~\cite{Manninen:2008mg} in central $Au+Au$ collisions and we have
used this value throughout at all centralities while the $\mu_S$ and 
$\mu_Q$ are set to zero.

Once the chemical potentials are fixed, there are
three free parameters characterizing our system: the temperature,
the strangeness under-saturation parameter $\gamma_S$ and the overall
normalization volume $V$.
The auxiliary parameter $\gamma_S$  is necessary to include in our
analysis in order to take into account the empirical fact that the
strange particle yields are strongly suppressed with respect to
SHM estimates in elementary particle collisions. We have chosen
$\gamma_S$=0.6 in our analysis in accordance with statistical
hadronization model fits~\cite{Adams:2003xp,Becattini:2010sk} to
proton-proton collisions at this beam energy. In $Au+Au$
collisions, the $\gamma_S$ parameter increases monotonically from
the $p+p$ value to unity as a function of centrality~\cite{Manninen:2008mg}
and the
effect will be discussed in detail in the forthcoming sections.
The common normalization factor $V$ for all hadron species
is determined in different collision systems from the measurements
(see below)
while for the temperature we have chosen the value
$T=170$ MeV in all systems based on the SHM fits to proton-proton and
$Au+Au$ data at this beam energy.

We have included the same collection of hadron species in our
analysis as has been included in the works quoted above from which
we have taken the thermal parameters. The mean primary hadron and
resonance yields of each of the hadron species included in the
analysis are calculated according to Eq. (\ref{dndy}).
For resonances with width larger than 2 MeV, Eq. (\ref{dndy}) is
convoluted with the relativistic Breit-Wigner distribution and
integration over the mass and momentum is enforced. Once the mean
primary yields are known, we assume that, from event-to-event,
the multiplicity distribution of each species is governed by the
Poisson distribution characterized by the mean multiplicities
evaluated with Eq.~(\ref{dndy}). We then sample the Poisson distributions
for each of the hadron species in each event to obtain the primary
multiplicities and choose momenta for every hadron according to the
Boltzmann distribution. All unstable resonances are then
allowed to decay according to the most recent branching fractions
taken from the particle data group tables~\cite{Amsler:2008zzb}.
This way we know the momenta of every final state particle and
thus it is straightforward to take into account any geometrical
and kinematical experimental cuts.

We note here that the statistical hadronization model is useful
only for evaluating the relative yields of different hadron
species since the rapidity and transverse momentum spectra of all
hadrons emitted in the high energy collision experiments do not
resemble thermal distributions. This is not a problem for us as
long as our results do not depend explicitly on the details of the
spectra. Indeed, this is the case for the ''true'' dilepton
emission from a single decaying hadron, i.e. the invariant mass of
the lepton pair does not depend on the momentum of the parent
hadron and neither does the (Lorentz invariant) number of hadrons.
Thus, we may evaluate the number of produced hadrons and
dileptons within the statistical hadronization model even though
the spectra are not correct.

Unfortunately, the discussion above holds true only if the
measurement is performed in $4\pi$, i.e. if all hadrons are
measured or at least if the experiment extrapolates the measured
kinematical region to the unmeasured regions as well. The PHENIX
dilepton measurement is carried out in a narrow rapidity window
around $y=0$ and only leptons with $p_T>$ 200 MeV are taken into
account. We have also taken into account the azimuthal geometry
and effects of the magnetic field on the charged leptons of
the PHENIX detector~\cite{Adare:2009qk}
, namely, that electrons and positrons (of charge $q$)
are accepted in case both of the conditions
\begin{eqnarray}
\phi_{\rm min} &\le& \phi + q\frac{k_{\rm DC}}{p_T} \le \phi_{\rm max}
\,\,\, ; \,\,\, k_{\rm DC}=0.206\,\, {\rm rad}\,\, {\rm  GeV/c}
\nonumber \\
\phi_{\rm min} &\le& \phi + q\frac{k_{\rm RICH}}{p_T} \le \phi_{\rm max}
\,\,\, ; \,\,\, k_{\rm RICH}=0.206\,\, {\rm rad}\,\, {\rm  GeV/c} \nonumber
\end{eqnarray}
are simultaneously fulfilled. The PHENIX detector consists of two arms
with the angular coverage
$\phi_{\rm min}=-\frac{3}{16}\pi$, $\phi_{\rm max}=\frac{5}{16}\pi$ and
$\phi_{\rm min}=\frac{11}{16}\pi$, $\phi_{\rm max}=\frac{19}{16}\pi$.

The limited acceptance in rapidity is not a severe
problem here due to the approximate boost invariance of the systems around
mid-rapidity at top RHIC energies. In order to make the rapidity 
distributions of
non-charmed hadrons wider, we have randomly
boosted\footnote{assuming a Gaussian distribution with width
$\sigma_y=4.2$} (event by
event) our ''fireball'' along the beam axis so that the rapidity
distributions of pions become compatible with the BRAHMS
measurements~\cite{Bearden:2004yx}.

\begin{figure}[!ht]
\includegraphics[angle=0,scale=1.0]{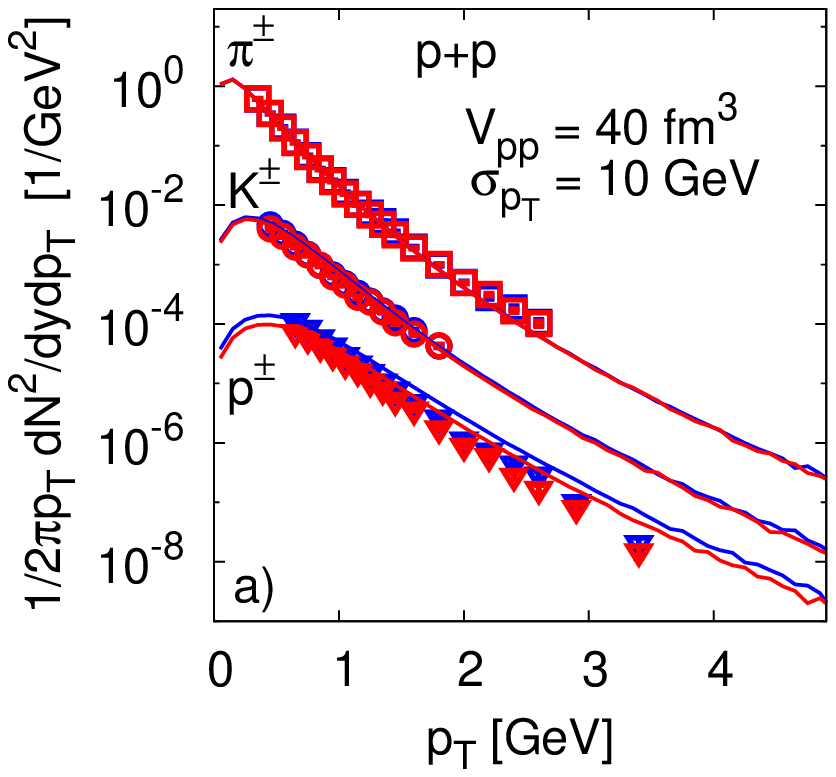}\\
\includegraphics[angle=0,scale=1.0]{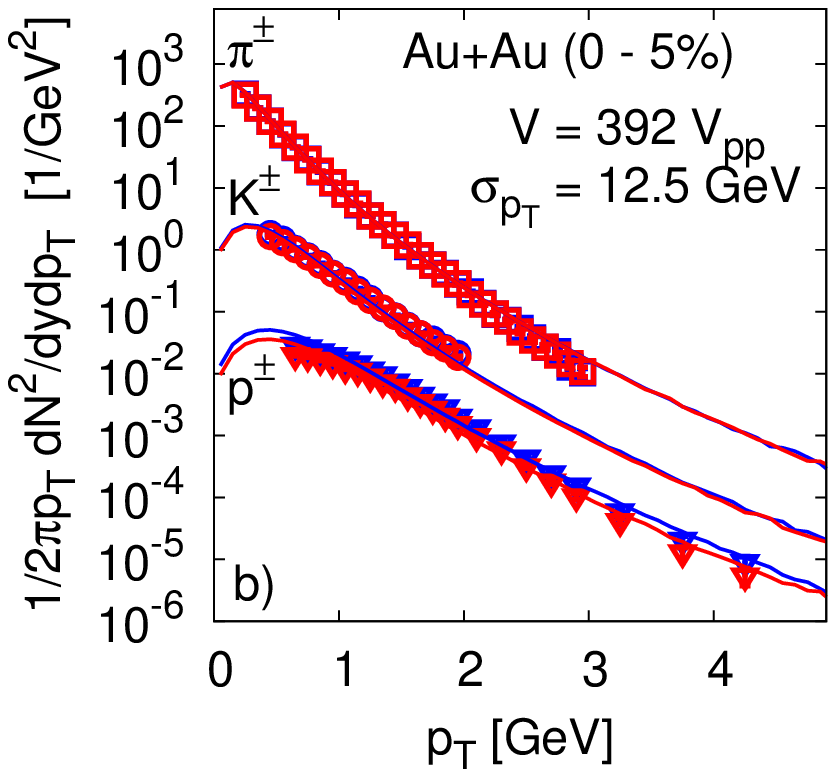}
\caption{(Color on-line) Transverse momentum spectrum of
$\pi^+$, $\pi^-$, $K^+$, $K^-$, $p$ and $\bar{p}$
~\cite{Adler:2006xd,Adler:2003cb} in
proton-proton collisions (top panel) and (0-5\%) most central
$Au+Au$ collisions (bottom panel) at \sqrtsnn=200 GeV compared
with our model calculations. The width of the clusters' transverse
flow profile along with the system volume are fitted to the data shown
in the figure. Kaons have been divided by a factor of 10 and nucleons by 
a factor of 100 for clarity.
\label{pTspectra}
}
\end{figure}

\begin{figure}[!ht]
\includegraphics[angle=0,scale=1.0]{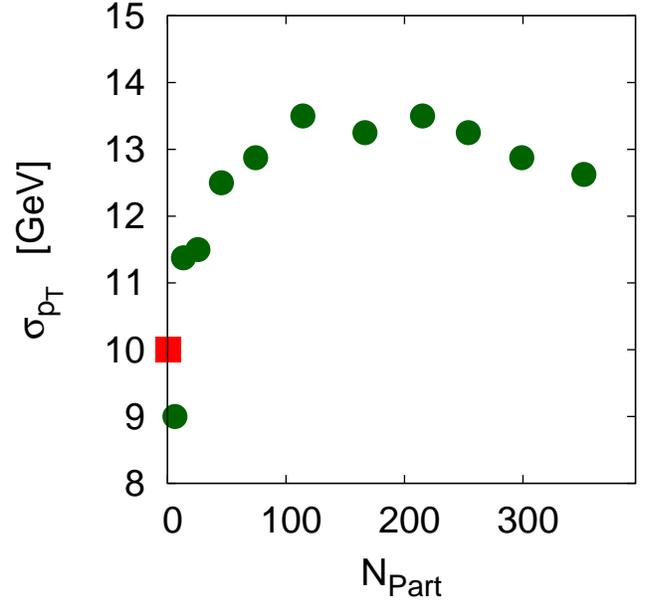}
\caption{(Color on-line) The width of the clusters' transverse
momentum distribution in 11 centrality classes in $Au+Au$ collisions
at \sqrtsnn=200 GeV (filled spherical symbols) as a function of the number
of participants. The square symbol denotes the width in $p+p$
collisions.
\label{sptwidth}
}
\end{figure}

\begin{figure}[!ht]
\includegraphics[angle=0,scale=1.0]{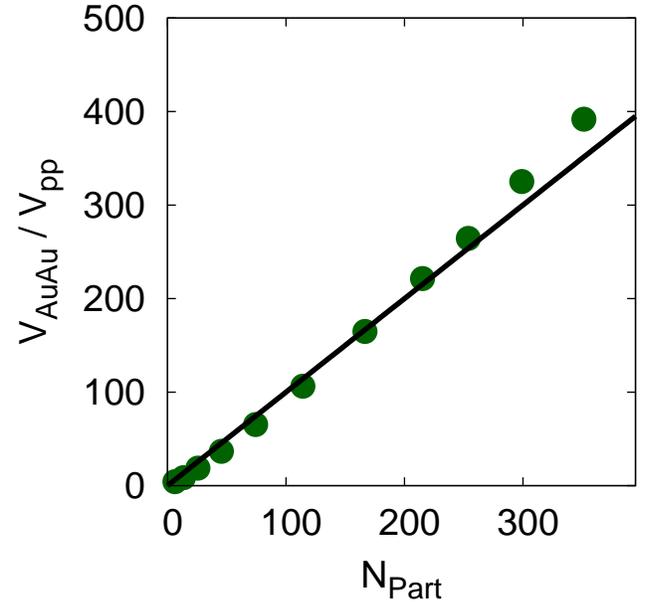}
\caption{(Color on-line)
The system volume normalized to the volume in $p+p$ collisions
in 11 centrality classes in $Au+Au$ collisions
at \sqrtsnn=200 GeV as a function of the number
of participants. Deviations from the $\NP$ scaling (straight line)
are visible only in the central collisions.
\label{volume}
}
\end{figure}

On the other hand, the limited acceptance in $p_T$ raises some
problems, because the statistical hadronization model tends to
over-populate the low $p_T$ part of the spectrum compared with the
experimental distributions and thus too few leptons hit the PHENIX
acceptance window of $p_T>$ 200 MeV. We have solved this problem by
assuming that the created clusters' transverse momentum is
normally distributed (with mean $\mu_{p_T}$=0 but $\langle p_T
\rangle > 0$) and fitted the width of the clusters' $p_T$
distribution together with the system volume $V$ to the PHENIX
data~\cite{Adler:2006xd,Adler:2003cb} in
$p+p$ collisions and in 11 different centrality classes in the case
of $Au+Au$ collisions. Comparison of the data and model calculations
are shown in Figure~\ref{pTspectra} for the $p+p$ (top panel)
and central $Au+Au$ collisions (bottom panel). The description of
the data is similar at all centralities. The resulting widths
for the clusters' transverse momentum distributions are shown in
Figure~\ref{sptwidth} while the scaling volumes (divided by the
volume in proton-proton collisions) are shown in Figure~\ref{volume}.

The dilepton yields are measured in wider centrality classes than
the $p_T$ spectra from which we have determined the $\sigma_{p_T}$
and $V$ and thus in the following calculations for dielectrons we
have used interpolated values for $\sigma_{p_T}$
and $V$ based on the values shown in Figures~\ref{sptwidth} and
\ref{volume}.

\section{Decay widths}

In the low invariant mass region the dominant sources of
(correlated) dileptons are the direct and Dalitz decays of light
mesons. The dielectron decay channels taken into account in this
analysis are listed in the Table \ref{dilcha}. Each of the direct
decays results in a sharp peak in the mass spectrum at the meson
nominal mass while the Dalitz decays yield a continuum spectrum
from zero invariant mass up to the mass of the decaying meson. Let
us note here that there are many other hadrons and their
resonances decaying radiatively into dileptons than the ones
listed in Table \ref{dilcha}. These could be (and are) important in
different kinds of collision systems. For example in heavy-ion
collisions at low beam energies, the Dalitz decay of $\Delta$
resonances dominate the low mass region of the dilepton invariant
mass spectrum (see e.g.~\cite{Wolf:1990ur,Schmidt:2008hm,lena2008}).
Above mid
SPS beam energies, however, the number of mesons exceed the number
of baryons  in heavy-ion collisions and at RHIC beam energies,
the emission of dileptons from baryons is overwhelmed by orders
of magnitude by the mesonic sources. Thus, we do not consider the
dileptons stemming from decays of baryons in this work because
the contribution is completely negligible at all invariant masses.

\begin{table}[!ht]
\begin{tabular}{|c|c|c|c|}
\hline
Hadron & direct & Dalitz & other \\
\hline
$\pi^0$ & - & $\pi^0\rightarrow \gamma\, \e^+\e^-$ & -\\
$\eta^0$ & - & $\eta^0\rightarrow \gamma\, \e^+\e^-$ &
$\eta^0 \rightarrow \pi^+\pi^- \e^+\e^-$\\
$\eta'$ & - & - &
$\eta' \rightarrow \pi^+\pi^- \e^+\e^-$\\
$\rho^0$ & $\rho^0\rightarrow \e^+\e^-$ & - & - \\
$\omega^0$ & $\omega^0\rightarrow \e^+\e^-$ & - &
$\omega^0\rightarrow \pi^0\, \e^+\e^-$ \\
$\phi^0$ & $\phi^0\rightarrow \e^+\e^-$ & - &
$\phi^0\rightarrow \eta\, \e^+\e^-$\\
$J/\psi$ & $J/\psi\rightarrow \e^+\e^-$ &
$J/\psi\rightarrow \gamma\, \e^+\e^-$ & -\\
$\psi'$ & $\psi'\rightarrow \e^+\e^-$ &
$\psi'\rightarrow \gamma \, \e^+\e^-$ & - \\
\hline
\D~mes. & - & - & $D^\pm \rightarrow \e^\pm \nu_e + X$\\
\hline
\end{tabular}
\caption{List of decay channels relevant for dielectron
production in $p+p$ collisions at $\sqrt{s}$=200 GeV. For $D^\pm$
mesons, 7 semileptonic (electron(positron) +
anti-neutrino(neutrino) + 1 or 2 light hadrons) decay channels are
considered. For the neutral ($D^0$ and $\bar{D^0}$) mesons and
$D_s^\pm$, there are 6 semileptonic decay channels taken
into account} \label{dilcha}
\end{table}

The decay probability of a meson into a pair of leptons depends on
the invariant mass of the lepton pair. A generic expression for
the decay probability is known from Ref.~\cite{Landsberg:1986fd}
\begin{eqnarray} \label{dgdm}
\frac{d\Gamma^{X \rightarrow \gamma \,\,
l^+l^-}}{dM} = \frac{\Gamma^{X \rightarrow l^+l^-}}{M}
\frac{4\alpha}{3\pi} \sqrt{1-\frac{4m_l^2}{M^2}} \nonumber\\
\times \Big(1+\frac{2m_l^2}{M^2} \Big)
\Big(1-\frac{M^2}{m_X^2}\Big)^3
|F^{X\rightarrow\gamma\gamma}(M)|^2.
\end{eqnarray}
Here $m_l$, $m_X$ and $M$ are the masses of the lepton, the decaying meson
and the
invariant mass of the dilepton pair, respectively. The form
factors $F^x(M)$ have been studied extensively both experimentally
and within different models. In this work we have employed the
form factors arising from the vector-meson dominance model
considerations~\cite{Cassing:1999es,Landsberg:1986fd}
\begin{eqnarray}
F^{\eta \rightarrow \gamma\gamma}(M) &=& \Big(1- \frac{M^2}{(0.72 {\rm GeV})^2}
\Big)^{-1} \nonumber \\
F^{\pi^0 \rightarrow \gamma\gamma}(M)&=& 1+ \frac{5.5}{\rm GeV^2} M^2.
\end{eqnarray}
The $\omega \rightarrow \pi^0l^+l^-$ channel is calculated from
\begin{eqnarray} \label{dgdm_omega}
&&\frac{d\Gamma^{\omega \rightarrow \pi^0 \,\,
l^+l^-}}{dM} = \frac{\Gamma^{\omega \rightarrow \pi^0\gamma}}{M}
\frac{2\alpha}{3\pi} \sqrt{1-\frac{4m_l^2}{M^2}}
\Big(1+\frac{2m_l^2}{M^2} \Big) \\
\times&&
\Big[
\Big(1+\frac{M^2}{m^2_\omega-m^2_\pi}\Big)^2
-\frac{4 m^2_\omega M^2}{(m^2_\omega - m^2_\pi)^2}
\Big]^{3/2}
|F^{\omega\rightarrow \pi^0 l^+l^-}(M)|^2 \nonumber
\end{eqnarray}
with the form factor
\begin{equation}
|F^{\omega \rightarrow \pi^0 l^+ l^-}(M)|^2 = \frac{(0.67 {\rm GeV})^4}
{((0.67 {\rm GeV})^2-M^2)^2 + (0.0516 {\rm GeV^2})^2}.
\end{equation}

The decay widths for the direct decays of vector mesons depend on
the mass of the decaying resonance. In practice this matters for
$\rho^0$ direct decays only, since all other mesons are
sufficiently narrow that the decay width can be considered
constant. For the $\rho^0$ direct decay, the decay width
reads~\cite{Bhaduri}
\begin{equation}
\Gamma^{V\rightarrow l^+l^-}(M)
= \frac{m_0^3}{M^3} \Gamma^{V\rightarrow l^+l^-}(m_0)
\label{rhowidth}
\end{equation}
in which $M$ is the $\rho$ -meson mass and $m_0$ denotes the
pole mass. The mass dependent branching fraction of the $\rho^0$
meson into a pair of leptons is obtained from Eq. (\ref{rhowidth})
by dividing it with the mass dependent total width of the
$\rho^0$ meson:
\begin{equation}
\Gamma^{\rho^0}_{tot}(M) \simeq \Gamma^{\rho\rightarrow\pi\pi}
= \Gamma(m_0) \Big( \frac{m_0}{M} \Big)^2
\frac{(M^2-4 m_\pi^2)^{3/2}}{(m_0^2-4 m_\pi^2)^{3/2}}.
\end{equation}

Above the $\phi$ meson mass, the dilepton invariant mass spectrum
attains contributions mainly from the decays of charmed hadrons.
Radiative decays of \jpsi~'s into dileptons have been studied in
detail in~\cite{Spiridonov:2004mp}. Since it is not possible to
(completely) disentangle the direct ($J/\psi \rightarrow l^+l^-$)
and Dalitz decays ($J/\psi \rightarrow \gamma\,\, l^+l^-$) of
\jpsi~in the collision experiments, one needs to take into account
also the Dalitz decay of \jpsi~in the analysis. This will modify
somewhat the shape of the invariant mass spectrum of the
dileptons stemming from decays of charmonia. We have implemented
the analytical formula~\cite{Spiridonov:2004mp}
\begin{eqnarray}
\frac{d\Gamma^{X\rightarrow l^+l^-\gamma}}{dM}
=\frac{\alpha}{\pi}\frac{2M}{m_X^2-M^2}
\Big(1+\frac{M^4}{m_X^4}\Big) \nonumber\\
\times\Big(\ln\frac{1+r}{1-r}-r\Big)\Gamma_0^{X \rightarrow l^+l^-} ,
 \label{jpsi}
\end{eqnarray}
which has been used successfully in describing the spectral
shape of the radiative decays of \jpsi~ measured both in DESY as
well as in PHENIX $p+p$ collisions~\cite{:2008asa}. In Eq.
(\ref{jpsi}) $m_X$ is the mass of the decaying particle, $M$ the
invariant mass of the dilepton pair and $r=\sqrt{1-4m_{l}/M^2}$.
This distribution diverges when $E_\gamma \rightarrow 0$ ($M
\rightarrow m_X$) and thus a cut in energy must be introduced. A
suitable value for the mass cut-off has been
found~\cite{Spiridonov:2004mp} to be around $E_\mathrm{min}
\approx$ 10 MeV, which we have also employed. Integrating Eq.
(\ref{jpsi}) gives us the widths of the radiative charmonia
decays: $\Gamma^{J/\psi\rightarrow \gamma\,\e^+\e^-} =
0.32\,\Gamma^{J/\psi\rightarrow \e^+\e^-}$ and
$\Gamma^{\psi'\rightarrow \gamma\,\e^+\e^-}$ = 0.34\, $
\Gamma^{\psi'\rightarrow \e^+\e^-}$. We have used these widths in
evaluating the branching ratios for the charmonia Dalitz decays in
our calculations. It is worth mentioning that our estimated
branching fraction BR(\jpsi$\rightarrow \gamma\,\e^+\e^-$) is
about twice the value listed in the most recent PDG
book~\cite{Amsler:2008zzb}. However, our choice agrees
somewhat better with the shape of the dielectron spectrum near the
\jpsi~peak than in the case of the PDG value for the \jpsi~Dalitz
decay branching ratio. The branching fraction for the $\psi'$
Dalitz decay is not yet measured and thus a comparison is not
possible.

For all other hadrons - not explicitly mentioned above -
we have assumed a relativistic Breit-Wigner spectral function and
the partial widths are then evaluated in a simplified procedure taking
into account only trivial mass threshold effects for the different decay
channels to correct for the available phase space.

\section{Charmonium and continuum background of dileptons from heavy quark decays}

The weak decays of \D~mesons $D \rightarrow l + \nu_l + h$, in
which $h$ denotes one or two non-charmed hadrons, constitute the main source
of the "dilepton continuum'' in the intermediate mass region at
RHIC energies. At RHIC energies in proton-proton collisions,
there is most often zero or a single charmed quark-anti-quark pair
created. When this pair of charmed quarks hadronizes, the most
likely result is that each of the charmed quarks end up in a
\D~and $\bar{D}$ meson. When the \D~mesons subsequently decay into
leptons and hadrons, we may have an extra lepton pair stemming
from the \D~meson decays in the final state. It is very difficult
to subtract the leptons originating from the \D~meson decays in the
collision experiments and thus the measured dilepton invariant
mass spectrum usually includes this contribution. The contribution
of the \D~meson decays is only significant far away from the true
sources of
dileptons and dominates the spectral shape between the 'peaks' of
the $\phi$ and \jpsi~mesons. Thus, one needs to carefully
consider the \D~mesons as  dilepton emitting sources in high-energy
collision experiments.

The slope of the dilepton continuum in the IMR arises as a
superposition of the momentum distributions of the measured
leptons coming from the individual \D~meson decays. Since the
measurement is carried out at mid-rapidity and we are interested
in invariant masses larger than 1 GeV, it is clear that the major
contribution to the invariant mass of the dilepton continuum in
the IMR arises from the transverse momenta of the decaying open
charm mesons. Accordingly, it is important to model the transverse
momentum spectrum of the \D~mesons more carefully than in the
longitudinal direction, i.e. the rapidity distribution.

We will employ a very simple model to evaluate the rapidity distributions
of charmed hadrons. Namely, we assume that all charmed quark-anti-quark
pairs are produced via splitting of a hard gluon created in the initial hard
collisions. We also assume that
the hard gluons - from which all $c\bar{c}$-pairs originate - are created
via gluon-gluon fusion processes in the initial hard scatterings
of the gluons from the target and projectile. In this case, the
final charmed hadron rapidity distributions will closely follow the
rapidity distribution of the hard gluons emitted in the collision
experiment.

In order to fix the rapidity distributions of charmed
hadrons we harness the idea of limiting
fragmentation~\cite{lim_frag}, which has been verified
experimentally at ultra-relativistic beam energies both in
hadron-hadron~\cite{Elias:1979cp,Alner:1986xu} as well as in
heavy-ion collisions~\cite{Back:2002wb}. Let us consider a
collision of two gluons with momentum fractions of $x_1$ and $x_2$
of the colliding projectile and target. One can show (see e.g.
~\cite{Stasto:2007zz}) that the resulting parton rapidity
distribution at large momentum fraction (i.e. $x_1 \gg x_2$ or
vice versa) is proportional to the parton distribution function
itself \begin{equation} \label{dndycharm} \frac{\d N}{\d y} \sim
x_1 g(x_1) \,\,\,\,\,\,\,\,\,\,\,\,\,\, \textrm{;}
\,\,\,\,\,\,\,\,\,\,\,\,\,\, x_1 =
\frac{p_T}{m_N}\e^{y-y_\mathrm{beam}} \end{equation} and is
approximately independent on the $Q^2$ scale due to Bjorken
scaling. In Eq. (\ref{dndycharm}) $p_T$ and $y$ are the transverse
momentum and rapidity of the produced gluon, $m_N$ is the mass of
the beam particle (in this case the proton). Thus, in order to
estimate the rapidity distribution of the charmed hadrons, we have
to adopt a suitable parametrization for the parton distribution
function in Eq. (\ref{dndycharm}). We have chosen the
following NNLO pQCD best fit parametrization from
Ref.~\cite{Alekhin:2005gq} for our gluon distribution
\begin{eqnarray}\label{xgx}
xg(x) \sim x^a(1-x)^b(1+\gamma_1\sqrt{x}+\gamma_2 x) \,\,\,\,\,\,
\textrm{with}\nonumber\\
a = -0.118 \,\,\,\, b = 9.6 \,\,\,\, \gamma_1=-3.83 \,\,\,\,
\textrm{and} \,\,\,\, \gamma_2 = 8.4.
\end{eqnarray}
This parametrization is given at $Q^2$=9 GeV$^2 \approx M_{{\rm J}/\psi}^2$
and we approximate the gluon distributions at higher $Q^2$ with the
same parametrization.

We can now evaluate the rapidity distribution of charmed hadrons with
a modified Brodsky-Gunion-Kuhn (BGK) model~\cite{Brodsky:1977de} introduced
in Ref.~\cite{Torrieri:2009fv}, in which the parton number density
of produced partons along the beam axis is proportional to a
triangle defined by the momentum fractions $x_1$ and $x_2$ as
follows: The center of mass of the colliding partons move with
rapidity $y_{cm}=\mathrm{atanh}(\frac{x_1-x_2}{x_1+x_2})$. We assume that
the probability along the rapidity axis to find the hard gluon -
fragmenting into a charmed quark pair - is defined by a triangle
whose maximum is at $y_{cm}$ and which goes linearly to zero at
$y=\mathrm{asinh}(x_1\sqrt{s}/2m_{\rm N})$ and
$y=-\mathrm{asinh}(x_2\sqrt{s}/2m_{\rm N})$.
The area of this triangle is set to unity so that it represents a proper probability.

\begin{figure}[!ht]
\includegraphics[angle=0,scale=0.88]{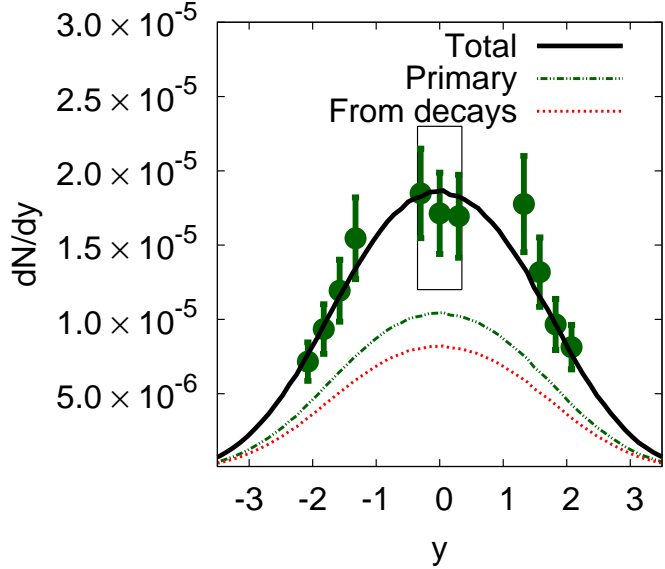}
\caption{(Color on-line) Calculated and measured \jpsi~rapi\-dity distribution
in proton-proton collisions at $\sqrt{s}$=200 GeV. The data
(full dots with error bars) are from the PHENIX
collaboration~\cite{Adare:2006kf}. The primary production of \jpsi,
$\chi_c$ and $\psi'$ are evaluated as described in the text. The
"From decays" -line denotes the feeding from $\psi'$ and $\chi_c$
states. The central rapidity region, in which the dilepton measurement is
carried out, is indicated by the vertical lines.
\label{fig:jpsi_y}}
\end{figure}

We have estimated the charmonium cross sections by the expression
(taken from Ref.~\cite{Linnyk:2008hp})
\begin{equation}
\sigma_i^{NN}(s) = f_i a \Big(1-\frac{m_i}{\sqrt{s}}\Big)^\alpha
\Big(\frac{\sqrt{s}}{m_i}\Big)^\beta
\theta(\sqrt{s}-\sqrt{s_{0i}})
\label{olenapara}
\end{equation}
in which $\sqrt{s}$ is the center-of-mass collision energy (per
nucleon pair) and $m_i$ is the mass of the charmonium state $i$.
The parameters $\alpha$=10, $\beta$=0.775 and $a$ are
common for all charmonia and fitted to experimental data. The
threshold factors read $\sqrt{s_{0i}}=m_i + 2m_ N$ while the
parameters $f_i$ were fitted separately in~\cite{Linnyk:2008hp}
for each of the states ($f_i$=0.636, 0.581 and 0.21 for $\chi_c$,
\jpsi~ and \psipr~, respectively).
Above, the multiplicity label
$\chi_c$ denotes the sum of the three $\chi_{0c}$,  $\chi_{1c}$
and  $\chi_{2c}$ states. All these states decay into \jpsi~ and we
have taken the sum of their branching ratios (0.55) into \jpsi~ as
our branching ratio for the generic "$\chi_c$". We have slightly
re-adjusted the common normalization factor $a$ from 0.16 mb to
0.133 mb in order to
reproduce exactly the total \jpsi~production cross section
$\sigma_{J/\psi}^{NN}=3.0\mu b$ measured
by the PHENIX collaboration~\cite{Adare:2006kf}.
The rapidity distribution of \jpsi~'s in proton-proton collisions
at $\sqrt{s}$=200 GeV, evaluated according to Eqs.
(\ref{dndycharm}), (\ref{xgx}) and (\ref{olenapara}) is compared
with the PHENIX measurement in Fig. \ref{fig:jpsi_y}.
The agreement appears good enough so that we can estimate both
the rapidity density at mid-rapidity as well as the total cross section
in our simple model.

The reason we have taken the trouble to set up a model that can describe
the \jpsi~rapidity distribution in $p+p$ collisions is that we need this
model to evaluate the rapidity distribution of \D~mesons. The open charm
rapidity distribution is not yet measured at this beam energy and thus we
need to calculate it. Since we have seen that our model can describe the
\jpsi~data in this collision system, we can fairly safely assume that the
same model will, at least approximately, describe the longitudinal part of
the open charm momentum distribution as well.

\begin{figure}[!ht]
\includegraphics[angle=0,scale=1.0]{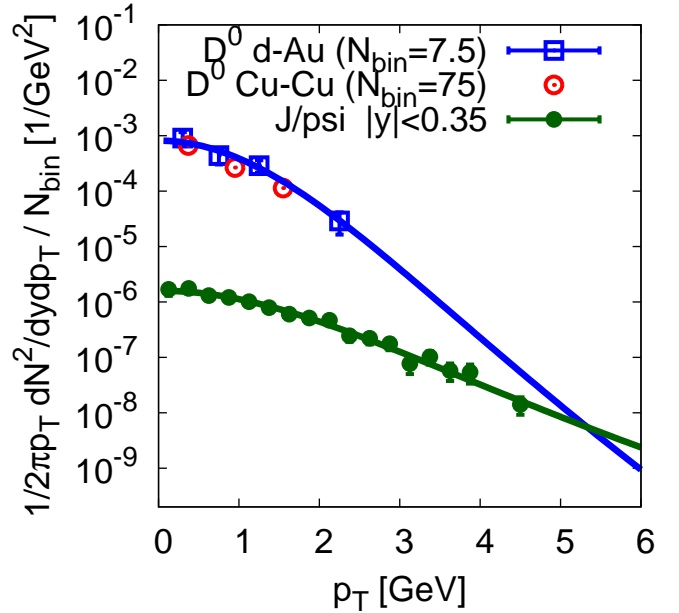}
\caption{(Color on-line) \D~meson $p_T$ spectra per binary collision in
$d+Au$~\cite{Adams:2004fc} and $Cu+Cu$~\cite{Lapointe} collisions as
well as the \jpsi~$p_T$ spectrum~\cite{Adare:2006kf} in $p+p$
collisions. The lines are power-law fits to the data as described
in the text.
\label{charmpTspectra}}
\end{figure}

Let us turn our attention to the transverse directions now. As we discussed
earlier, the transverse direction contributes most to the invariant mass
of the dileptons from open charm decays at mid-rapidity. This is why we
will rely on experimental data here.
The transverse momentum distributions of \D~mesons are
experimentally not well known, though. What is much better known is the
rapidity and transverse mass distribution of \jpsi~'s in
proton-proton collisions at RHIC. We deem that the momentum
distributions of \D~mesons resemble the corresponding ones for
\jpsi~ since the shape of the distribution - especially in the
beam direction - should be primarily determined by the dynamics of
the hadronizing charmed quarks. Thus, we assume that the \D~meson
transverse momentum distribution has a similar form as that for
\jpsi~'s in the same collision system. The $p_T$ spectrum of
\jpsi~ \cite{Adare:2006kf,Adare:2009js} (measured by the PHENIX
collaboration) can be described well with the power-law function
\begin{equation}
\label{charmdndpt}
\frac{dN}{dp_T} \sim (1+(p_T/B)^2)^{n},
\end{equation}
see Figure \ref{charmpTspectra}.
Here we have used the published data and fitted $B$=3.74 and $n$=5.11.

According to our best knowledge, the transverse momentum
distribution of \D~mesons have not been measured in proton-proton
collisions at RHIC beam energies. Preliminary data~\cite{Lapointe}
in $Cu+Cu$ collisions do exist as well as already published
data~\cite{Adams:2004fc,Tai:2004bf} in $d+Au$ collisions at
\sqrtsnn=200 GeV. The transverse momentum spectra of \D~mesons in
these collision systems - divided by the number of binary collisions -
along the corresponding \jpsi~data in $p+p$ collisions are shown in
Fig. \ref{charmpTspectra}. The lines shown have the
functional from of Eq. (\ref{charmdndpt}) and are fitted to the
data. The \D~meson data do not allow to reliably fit both $B$
and $n$ (as well as the normalization) and so we have chosen to
fix the parameter $B=3.74$ as in the case of \jpsi~ and re-fitted
$n=10.7$ in order to describe the STAR data for $d+Au$ collisions.

The dilepton continuum stemming from \D~meson decays attains an
extra feature compared with the dilepton emission from other
sources. Namely, since the electron and positron are emitted by
two different hadrons, the angular correlation as well as
re-scattering effects can strongly alter the invariant mass
spectrum of the final state dileptons. We assume here that the emitted
leptons themselves always escape the collision zone unscathed and
(re-)scattering effects can only take place on the hadronic level.
The angular correlations of the open charm hadrons are
experimentally not well known and thus we will employ
theoretical estimates.

We deem that the two charmed quarks are always emitted in
180$\,^{\circ}$ angle in their respective center-of-mass frame while
this angle is typically much smaller in the laboratory frame due to
large Lorentz boosts in the longitudinal direction, 
especially at large forward and backward rapidities. We
assume here that the longitudinal direction is not different from the
transverse directions in the CM frame of the fragmenting $c\bar{c}$-pair, i.e.
that Eq. (\ref{charmdndpt}) describes the joint distribution of any two
momentum components "$p_T$"=$\sqrt{p_x^2 + p_y^2}$=
$\sqrt{p_x^2 + p_z^2}$=$\sqrt{p_y^2 + p_z^2}$ in the CM frame of the
charmed hadrons. The angular distribution among the produced \D~mesons
is then taken exactly back-to-back correlated in their respective CM frame
and we need to boost the momenta of the produced \D~mesons into the
laboratory frame in order to evaluate the angular correlations in that particular frame.
We have cross-checked our approach and verified that our
angular correlations - evaluated as described above -
agree well with correlations evaluated with
the PYTHIA~\cite{Sjostrand:2006za} event generator. Alternatively,
we might also have adopted the angular correlations from the
PYTHIA simulations.

We have taken into account the 12 lightest \D~meson states
($D^\pm$, $D^{*\pm}$, $D^0$, $\bar{D^0}$, $D^{*0}$, $\bar{D^{*0}}$,
$D_s^\pm$ and $D_s^{*\pm}$) all of whose mass is around 2 GeV. We
have assumed that the relative primary multiplicity of these 12 states is
determined purely by their spin -degeneracy while the total number
of them is taken from the parametrization of Ref.
\cite{Linnyk:2008hp}. On top of this, we have taken into account
the empirical fact that in a jet fragmentation process,
hadrons that include a strange or anti-strange quark suffer
further suppression. We have used the canonical value 0.3 for such
a strangeness suppression factor for the $D_s$ and $D_s^*$ states
in our analysis as in Ref. \cite{Linnyk:2008hp} which stems from PYTHIA
calculations. One should not
confuse this factor with the $\gamma_S$ parameter included in the
thermal model analysis. The factor 0.3 here concerns the hard
scatterings only, while the $\gamma_S$ parameter takes into
account also some of the soft physics on top of the suppression on
the hard scattering level.

The kinematics of the decays of the excited \D~mesons can be
neglected because of the small mass difference between the $D^*$
and \D~states. In each of the cases, the excited state relaxes
itself by emitting a very soft pion  and thus the daughter \D~inherits practically
the momentum of the decaying parent state. We deem that by
properly considering all of the lightest \D~mesons (and their
relative abundances), we can then hope to extract the total charm
production cross section in our analysis. Alternatively we could
have just included the lowest lying \D~meson states, since these
are the only ones decaying into leptons, but in this case the
normalization factor, the number of final state
\D~mesons would not have a clear physical interpretation as a cross
section. Rather than taking the rapidity densities of different
charmed hadrons at mid-rapidity as free parameters, we wish to
estimate the total cross sections of the different states and
suitably distribute the produced charmed hadrons at different
rapidities. In order to estimate the rapidity density of \D~mesons
around mid-rapidity, we take the total cross section for open
charm
production\footnote{$\sigma_{c\bar{c}}$=$\sigma_{D^+}$+$\sigma_{D^0}$+
$\sigma_{D_s}$+$\sigma_{D^{*+}}$+$\sigma_{D^{*0}}$+$\sigma_{D_s^{*}}$=485 $\mu b$}
from the parametrization given in~\cite{Linnyk:2008hp} and distribute
these along rapidity as described earlier.

\section{Results for dileptons}

\subsection{Proton-proton collisions}

\begin{figure}[!ht]
\includegraphics[angle=0,scale=1.0]{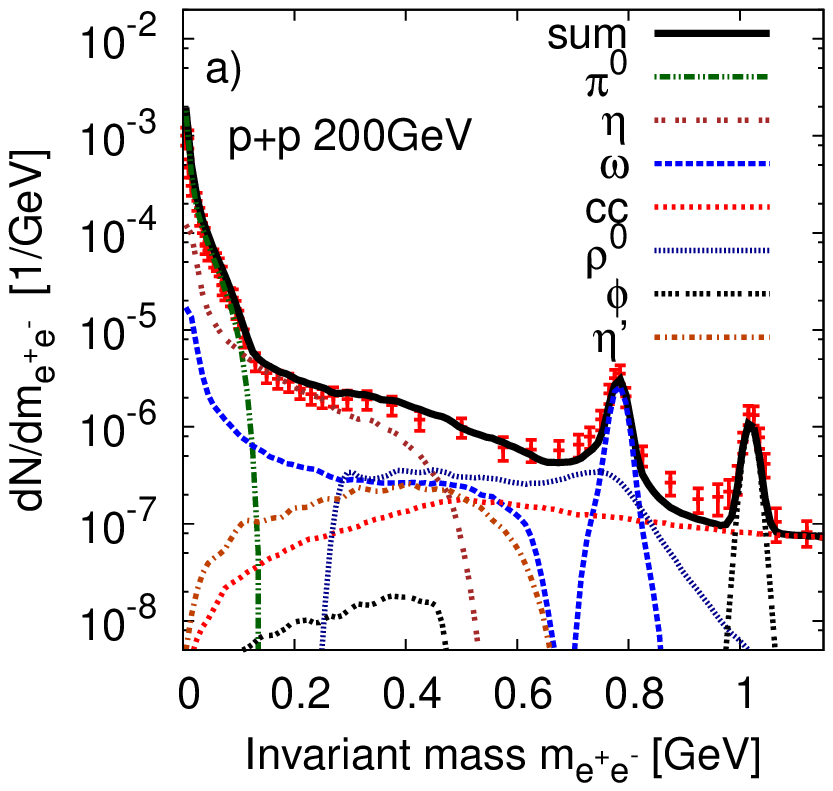}
\includegraphics[angle=0,scale=1.0]{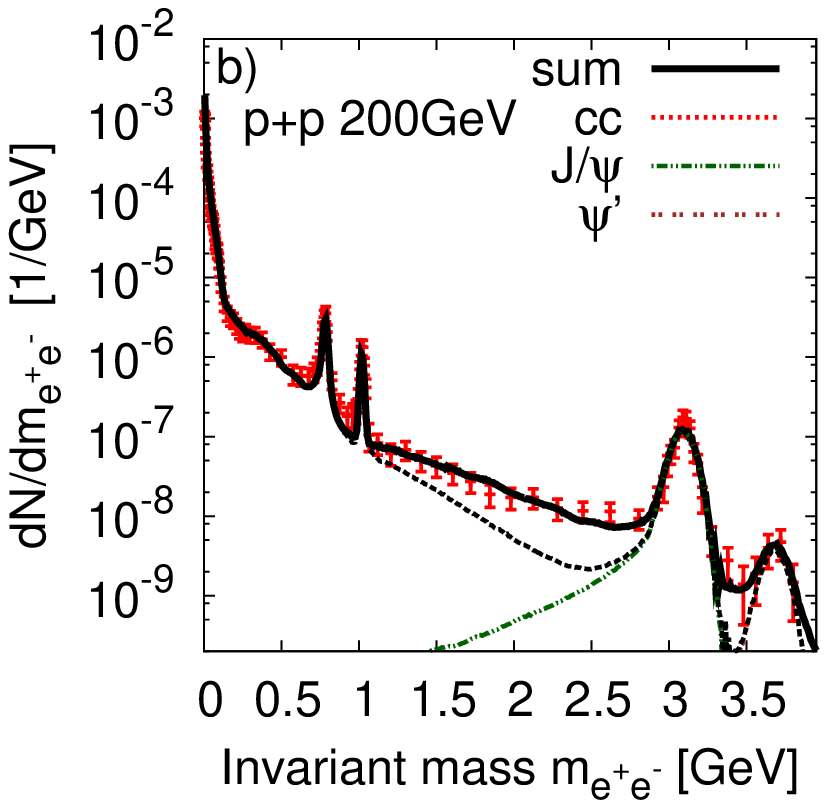}
\caption{(Color on-line) Invariant mass spectrum of pairs of electrons and
positrons in proton-proton collisions at $\sqrt{s}$ = 200 GeV. The
data are from the PHENIX collaboration~\cite{:2008asa} while the
contribution from different dilepton emitting sources are
calculated as explained in the text. The full thick black line
denotes the sum from all relevant sources. The LMR is shown in the
upper panel while the whole invariant mass range is shown in the
bottom panel. \label{fig:phenix-klein}}
\end{figure}

Let us first look at the proton-proton collisions. Our calculated
dielectron yields are compared with the experimental data in Fig.
\ref{fig:phenix-klein}. The top panel of Figure
\ref{fig:phenix-klein} shows the low mass region where the
emission is dominated by the decays of light mesons. One can see
that the spectrum can be reproduced very well within the
statistical hadronization ansatz described in the previous sections.
We mention that the $p+p$ dilepton mass spectra are also
well reproduced within the HSD transport approach \cite{lena2009}
where the charm production and angular correlations have been evaluated
within PYTHIA.

The shapes of the two peaks arising from the direct decays of
$\omega$ and $\phi$ mesons are essentially determined by the
experimental mass resolution. The natural width of these peaks
would be only about 8 and 4 MeV, respectively. The mass resolution
of the experiments, however, exceeds the vacuum width of these
hadrons and thus we have taken this into account by smearing the
peaks according to Gaussian distributions with a width
corresponding to an experimental mass resolution of
10 MeV for the light mesons except for the $\phi$ -meson direct decay 
whose resolution was taken to be 8.1 MeV~\cite{Seto:2004uy}. The mass
resolution of dielectrons from the charmed hadron decays was taken 
to be  2\% of $m_{e^+e^-}$.  By
doing so, especially the shape of the $\phi$ meson peak is
strongly modified and becomes compatible with the experimental
data as seen in Fig. \ref{fig:phenix-klein}. We have not
considered  in-medium modifications of the spectral functions
in this work and thus any apparent change in the vacuum spectral
functions is solely due to experimental acceptance cuts.

We find no
discrepancy\footnote{The invariant mass region between
the $\omega$ and $\phi$ meson peaks is somewhat under-estimated
in our approach. We deem that this is due to our approximation that
the angular correlations among the $D$ mesons are back-to-back
in the center-of-mass frame. In reality, a backward peaked but
narrow angular distribution is expected, which softens the spectrum
somewhat and populates the low invariant mass region. This effect,
however, would be only visible in the region between the $\omega$
and $\phi$ mesons.}
between the data and our model in the low
invariant mass region and can conclude that the data are very well
described within the statistical hadronization model assuming a
thermalized fireball. The only
non-equilibrium feature we have taken into account here is the
strangeness suppression factor $\gamma_S$. By this
factor we introduce a new - partly free but correlated -
normalization for hadrons consisting of one or more strange
quarks. In practice, the action of the $\gamma_S$ parameter is
best visible in the $\phi$ meson peak. Since there are no
resonances decaying into the $\phi$ meson the sole rapidity
density of the $\phi$ meson is calculated according to Eq.
(\ref{dndy}). With our choice for the value of $\gamma_S$ the
$\phi$ meson rapidity density is multiplied by a factor
$\gamma_S^2$=0.36. Obviously, without this extra strangeness
suppression the production of dileptons from the decays of $\phi$
mesons would be dramatically over-estimated. The $\gamma_S$
parameter affects also the yields of $\eta$ and $\eta$' mesons,
whose primary thermal production rates are multiplied by
$\gamma_S$ in order to take into account the fact that these
mesons are considered to carry ''hidden strangeness''. Thus the
value of $\gamma_S$ is mostly (but not solely) determined by the
$\phi$ meson. We remind the reader here that, besides the overall
normalization, we have not fine
tuned any of the thermal parameters in this work, instead, we have
used the values fitted in Ref. \cite{Becattini:2010sk} to STAR
data.

The whole invariant mass spectrum is shown in the lower panel of
Figure \ref{fig:phenix-klein}. Let us look at the IMR between the
$\phi$ and \jpsi~ peaks now. The solid and dashed black lines are
evaluated as explained in the last Section implementing the
transverse momentum profile fitted to the \D~meson transverse
momentum spectrum in $d+Au$ collisions. The only difference
between these two lines is that the upper solid line is evaluated
assuming back-to-back angular correlations in the center-of-mass
frame between the two fragmenting \D~mesons while the lower dashed line
is the same with random correlations.
From the figure it is clear that the model with random
correlations does not describe the IMR spectrum properly for $p+p$ reactions.

From this we can conclude, in accordance with the original PHENIX
publication~\cite{:2008asa}, that a
model with strong correlations (no final state interactions) among
the produced \D~mesons seems to be favored by the data over the
random correlation in case of proton-proton collisions.
We mention that in the extreme case of exact back-to-back correlations
in the laboratory frame among
the \D~mesons, the slope of the dilepton continuum in the IMR is
reproduced but in this case too few dileptons are emitted in the
LMR. We do not explicitly show these results but note that the
continuum between the $\omega$ and $\phi$ would be under-estimated
by about an order of magnitude in this over-simplified scenario.

We may conclude that once all relevant kinematical as
well as acceptance effects are taken into account, the
electron-positron invariant mass spectrum can be understood very
well in both  the LMR and IMR in proton-proton collisions within
our simple model and parametrization. This will serve as a
baseline for our comparative analysis in heavy-ion collisions for
different centralities. We mention in passing  that the
dilepton mass spectra from our simplified model agree with those
from the HSD approach \cite{lena2009} on the 20\% level when using vacuum spectral
functions for the hadrons in the transport approach.

\subsection{Heavy-ion collisions}

Let us turn  to heavy-ion collisions now. We have seen in the
previous Section that the dilepton invariant mass spectrum can be
well understood in proton-proton collisions and we will now try to
extrapolate our results from $p+p$ collisions to heavy-ion
collisions to identify the magnitude of possible additional production channels
from partonic sources as suggested e.g. in Ref. \cite{Linnyk:2009a} for SPS energies.

The non-charmed hadron yields are expected to scale with the
participant number ($\NP$) when comparing $p+p$ and heavy-ion
collisions while -  assuming that the charmed quark pairs are
created solely in the initial hard scatterings -  the number of binary
collisions ($\Nbin$) should be the correct scaling variable for
the charm sector. Both $\NP$ and $\Nbin$ can be estimated within the
Glauber model and are conventionally used also in the experimental
analyses. We have used the same values for $\NP$ and
$\Nbin$~\cite{Adler:2003cb,Abelev:2009tp} that are employed
by the experiments at RHIC in order to be consistent with similar
previous analyses.

\subsubsection{Low invariant mass region}

The $Au+Au$ and $Cu+Cu$ collisions are treated different in this paper.
For the $Au+Au$ collisions we have fitted the width of the clusters'
transverse momentum distribution and the system volume to the PHENIX data,
see Figures \ref{sptwidth} and \ref{volume}, and evaluated the dilepton
invariant mass spectrum in $Au+Au$ collisions implementing the 'optimized' 
parameters.
Detailed measurements of identified hadron transverse momentum spectra
are not yet available in the $Cu+Cu$ collisions and thus we have evaluated
the yields of dielectrons in the LMR in $Cu+Cu$ collisions such that we
use the transverse momentum profile fitted to the $Au+Au$ collisions
also in $Cu+Cu$ collisions at the same $\NP$. The volume in $Cu+Cu$
collisions at different centralities is evaluated by scaling the system
volume in $p+p$ collisions with the number of participants. As one 
can see from the
Figure~\ref{volume}, the fitted volumes scale with $\NP$
in the $Au+Au$ collisions in the $\NP$ region relevant for $Cu+Cu$ collisions
and thus we deem that the $\NP$ scaling should be a good approximation in
the $Cu+Cu$ case, at least in the non-central collisions.

\begin{figure}[!ht]
\includegraphics[angle=0,scale=1.0]{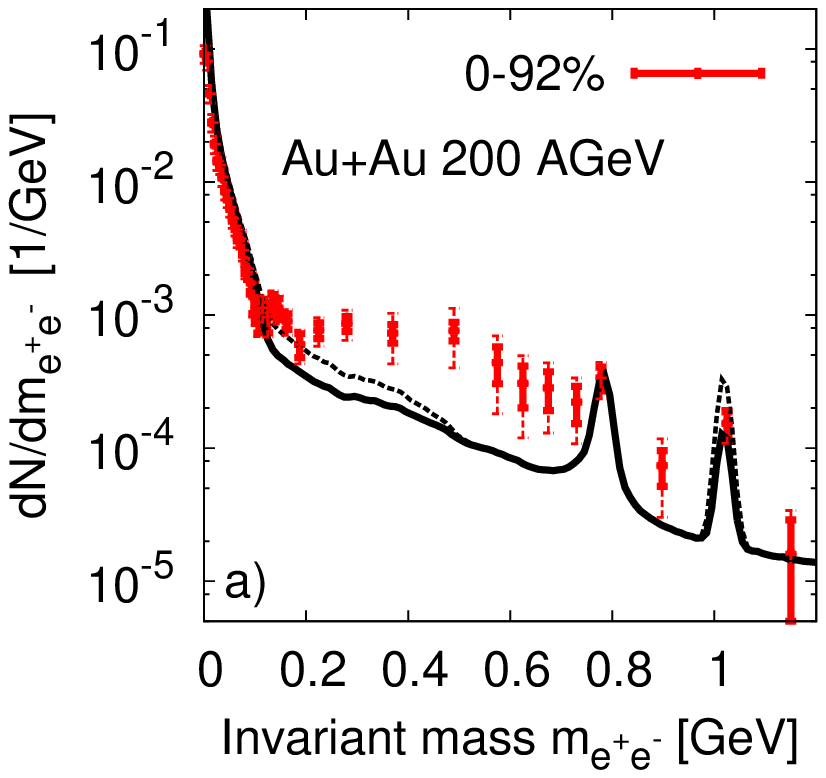}
\includegraphics[angle=0,scale=1.0]{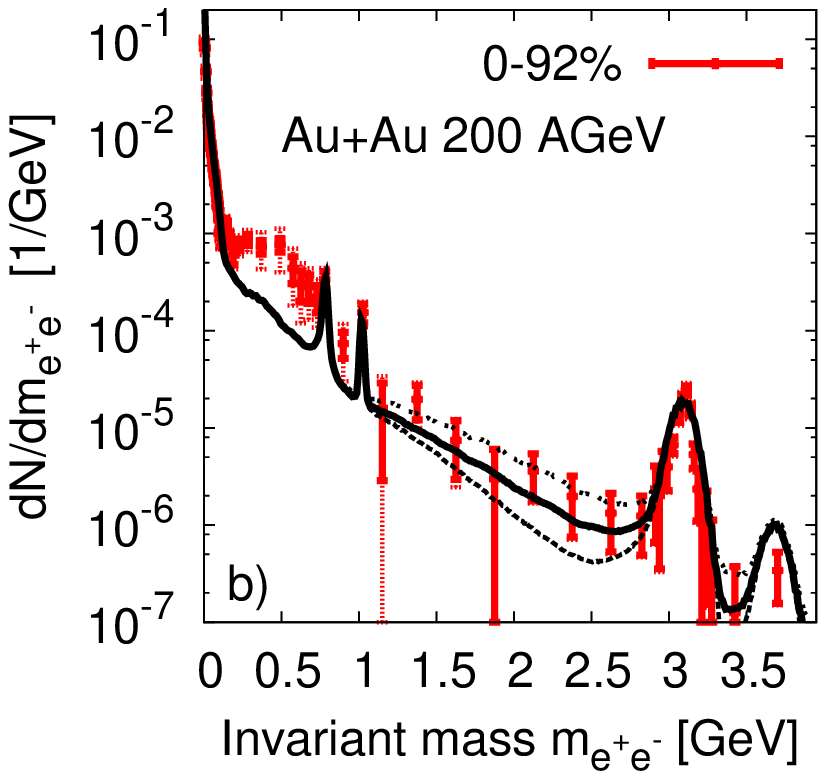}
\caption{(Color on-line) Invariant mass spectrum of pairs of
electrons and positrons in minimum bias $Au+Au$ collisions at
\sqrtsnn=200 GeV~\cite{:2007xw} compared with model calculations.
The low invariant mass region is shown in the top panel while the
results in the full invariant mass region are shown in the bottom
panel. The solid line indicates the model results scaled from
$p+p$ collisions such that the charmed hadron yields are scaled
with the number of binary collisions $N_\mathrm{bin}$ while the non-charmed 
hadron yields are scaled with 110 $V_{\rm pp}$ (the volume in $p+p$ collisions). 
The dashed curve (bottom panel) indicates the
results scaled from $p+p$ collisions assuming a random correlation
among the \D~mesons. The calculation with the $\gamma_S$ parameter
set to unity is indicated by the dashed line (top panel) slightly
above the solid one in the LMR. The dotted curve (bottom panel)
indicates the results evaluated with angular correlations kept the
same as in $p+p$ collisions. The solid and dotted error bars stand
for the statistical and statistical + systematic errors added in
quadrature, respectively. \label{fig7}}
\end{figure}

\begin{figure}[!ht]
\includegraphics[angle=0,scale=0.96]{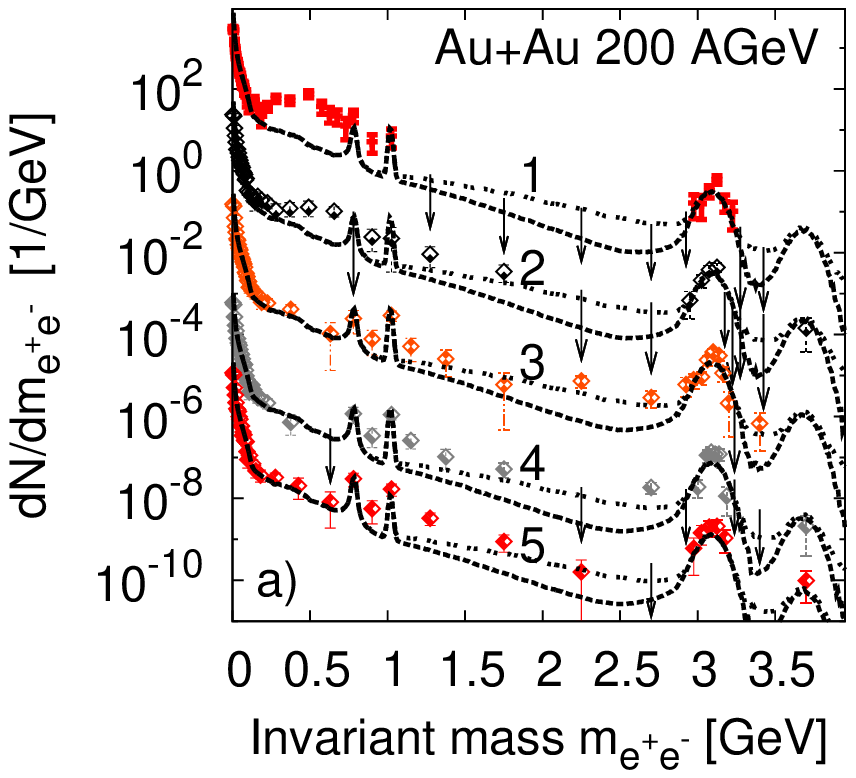}
\includegraphics[angle=0,scale=0.96]{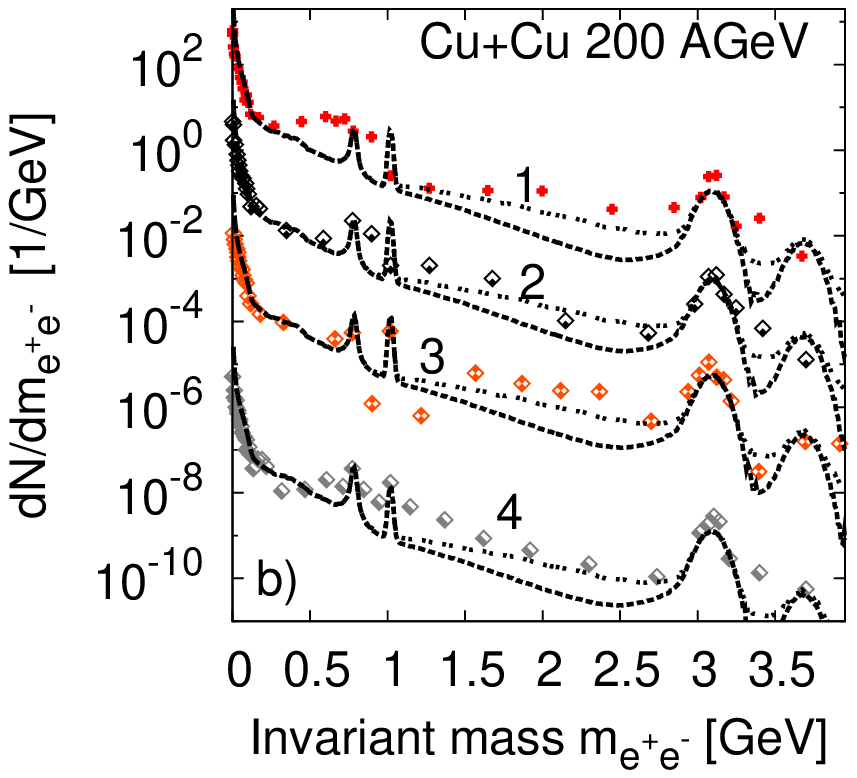}
\caption{(Color on-line) Invariant mass spectrum of pairs of electrons and
positrons in $Au+Au$ (top)~\cite{Adare:2009qk} and
$Cu+Cu$ (bottom)~\cite{Campbell} collisions at \sqrtsnn=200 GeV in
different centrality classes compared with the model calculations.
The centrality bins are labeled from central to peripheral as
1 (0-10\%), 2 (10-20\%),
3 (20-40\%). Centrality bin 4 consist on (40-94\%) and (40-60\%) most central
events in $Cu+Cu$ and $Au+Au$ collisions, respectively, while the centrality
bin 5 includes (60-92\%) most central collisions.
Both the data and model calculational results are scaled with factors of
$10^4$ (0-10\%), $10^2$ (10-20\%),
$1$ (20-40\%), $10^{-2}$ (40-60\%) as well as $10^{-3}$ (60-92\%) and (40-94\%)
for clarity. The double dotted lines indicate the model results scaled
from $p+p$ collisions such
that the charmed hadron yields are scaled with $N_\mathrm{bin}$. 
The non-charmed hadron yields are scaled with $\NP$ in the case of $Cu+Cu$
collisions while in $Au+Au$ collisions we have scaled the yields with 
the fitted volumes.
The dashed lines indicate the results scaled from $p+p$ collisions with random
correlations among the open charm hadrons. The starting point of the downward
pointing arrows denote the data points which are defined as upper limits only.
\label{fig6}}
\end{figure}

We have evaluated the dilepton yields in $Au+Au$ and $Cu+Cu$ collisions as
described above, see Figs. \ref{fig7} and \ref{fig6}. In the lower
panel of Fig. \ref{fig7} we show our results together with the experimental
data~\cite{:2007xw} in minimum bias $Au+Au$ collisions
in the whole invariant mass range while in the top panel a zoom
to the LMR is presented. In Fig. \ref{fig6} the
experimental results from Refs. \cite{Adare:2009qk,Campbell} in
different centrality bins both in $Au+Au$ (top panel) and
$Cu+Cu$ (bottom panel) collisions are shown. The $Cu+Cu$ data in different
centrality bins are still preliminary and the error bars are not
yet available.


In the following we will concentrate on discussing the $Au+Au$
collisions. Essentially the same conclusions will, however, hold
also for the $Cu+Cu$ collisions. Let us take a closer look at the
LMR first. From the top panel of Fig. \ref{fig6} one can see
that the data can be described well in the most peripheral
centrality bin in $Au+Au$ collisions in the LMR.
This centrality class is special among the centrality classes in
$Au+Au$ collisions because in all other centrality classes the
relative strangeness production is found to be nearly in chemical
equilibrium with $\gamma_S\approx1$ ~\cite{Manninen:2008mg}. We
have taken the increase in relative strangeness production into
account in our analysis by setting the $\gamma_S$ parameter to
unity in every other centrality bins except in the most peripheral
collisions, i.e. in the centrality bin labeled with ''5'' in Fig.
\ref{fig6}a and ''4'' in Fig. \ref{fig6}b. The results with both
$\gamma_S=0.6$ (solid) as well as $\gamma_S=1$ (dashed) are shown
for the minimum bias $Au+Au$ collisions in the top panel of Figure
\ref{fig7}. One can see that the increase in relative strangeness
production as a function of centrality is not strong enough to
explain the excess in the LMR in minimum bias $Au+Au$ collisions.

From Figure \ref{fig6}, one can see that the LMR data in $Au+Au$
collisions in peripheral and in semi-central bins can be
reasonably well described within the statistical hadronization
model. On the other
hand, in the two most central bins as well as in the minimum bias
collisions the increase in strangeness production can not explain
the excess of dileptons in the low invariant mass region from 0.15
to 0.6 GeV. In $Cu+Cu$ collisions, it seems that there is
significant excess in the LMR over the hadronic cocktail only in
the most central collision bin while the LMR is fairly well
described for the other centralities.

\begin{figure}[!ht]
\includegraphics[angle=0,scale=1.0]{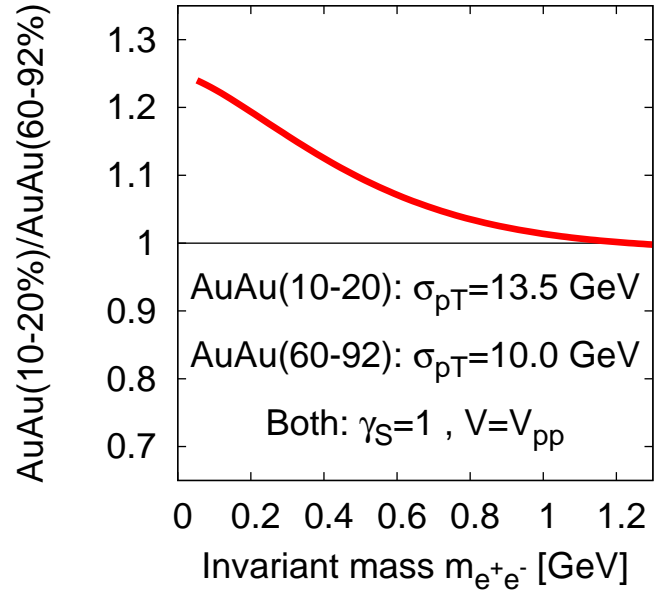}
\caption{(Color on-line) Ratio of invariant mass spectrum of electron-positron
pairs within the PHENIX acceptance evaluated with two different transverse flow 
profiles keeping other parameters fixed. 
\label{flowcomparison}}
\end{figure}

We have, furthermore,  studied the effect of the transverse flow on the dielectron
invariant mass spectrum in $Au+Au$ collisions by comparing the results
evaluated with the maximum and minimum width for the clusters'
transverse momentum distribution. The largest (see Fig.~\ref{sptwidth})
width, $\sigma_{p_T}$=13.5 GeV, was fitted in (15-20\%) most central
collisions while for the $p+p$ and (60-92\%) most central collisions
we use $\sigma_{p_T}$=10.0 GeV. We have evaluated the dielectron
invariant mass spectrum with these two widths keeping all their
parameters fixed ($T=170$ MeV, $V=V_{pp}$ and $\gamma_S=1$) and calculated
the ratio of the resulting invariant mass spectra of dielectrons in the
PHENIX acceptance, see Figure~\ref{flowcomparison}.

As expected, the
increase in the transverse flow enhances the dielectron yields in the
PHENIX acceptance especially at very low invariant masses while above
the $\omega$ meson peak equal amounts of dileptons hit the PHENIX acceptance
with both transverse flow profiles. In general the effect is moderate and the
increase in flow can increase the dielectron yields up to 30\% in the $\pi^0$
region, much less than the observed excess by factors up to 4-5.
Thus, one can conclude that the broadening of transverse momentum distributions
as a function of centrality plays only a minor role in the dielectron radiation
from the decays of light hadrons.

\subsubsection{Further correlated sources in the LMR}

Besides the decays of light mesons, there are other correlated sources
for dielectron production in high-energy nuclear collisions.
In this Section, we will address some but not all
of such processes. The contribution from most of these channels
is very small and thus we concentrate here on the dominant channels, only.
These additional channels, similarly to the
the \D~meson case, arise in correlated decays of light hadrons that do not
directly decay into electrons but produce single electrons via intermediate
hadrons. An example of such a process is the $\phi$-meson
decay $\phi \rightarrow K^+ K^-$. The $\phi$ meson has a long lifetime and
a small hadronic cross section and thus many of the $\phi$ meson decays
take place outside the fireball. Now it can happen that the kaons from
the $\phi$ meson undergo semileptonic decays like
$K^+ \rightarrow \pi^0 e^+ \nu_e$ and charge conjugate for the $K^-$. If
the $\phi$ meson decays outside the fireball, the correlations are
preserved and the experiment measures essentially additional correlated
dielectron radiation from the $\phi$ meson which survives the
experimental subtraction procedure from uncorrelated $e^+ e^-$ decays.

Unlike the $\phi$ meson, most of the short living
resonances are expected to decay inside the fireball thus destroying the
correlated signal. However, some of the interactions do take place in the
dilute corona of the fireball in which case the correlated signal can be
preserved even in central heavy-ion collisions. We do not perform a precise
calculation for these correlations and make an estimate
for the relative magnitude of core and corona emission, instead. Our
results thus have to be considered as an upper limit estimate for a correlated
background emission.

We have considered this type of correlated dielectron radiation from a
selection of fairly light strange and neutral hadrons whose branching
fractions into $K\bar{K}$ are sizeable and reasonably well known.
The additional channels we have studied in this work
are listed in Table~\ref{adddilcha}. We consider the kaonic channels only
in this work, even though most of the processes could proceed via the $\pi\pi$
channel as well. The pionic channels are found to be sub-leading compared
to the kaonic channels and thus we have omitted pionic channels in this work.

\begin{table}[!ht]
\begin{tabular}{|c|c|c|c|}
\hline
Hadron & & & \\
\hline
$f_0(980)$ & \KKbar & \KKbarnul & - \\
$f_1(1285)$ & - & - & $K\bar{K}\pi$  \\
$f_2(1270)$ & \KKbar & \KKbarnul & -  \\
\hline
$f'_0(1350)$ & \KKbar & \KKbarnul & -  \\
$f'_1(1420)$ & $K^{*+}K^-$ + c.c & $K^{*0}K^0$ + c.c & -  \\
$f'_2(1525)$ & \KKbar & \KKbarnul & -  \\
\hline
$f_0(1500)$ & \KKbar & \KKbarnul & -  \\
$f_1(1510)$ & $K^{*+}K^-$ + c.c & $K^{*0}K^0$ + c.c & -  \\
$f_2(1430)$ & \KKbar & \KKbarnul & - \\
\hline
$\phi$ & $K^+K^-$ & \KKbarnul & - \\
$a^0_0(980)$ & \KKbar & \KKbarnul & - \\
$K(892)^\pm$ & $K^\pm \pi^0$ & -  & - \\
$K(892)^0$ & $K^0\pi^0$ & - & - \\
\hline
\end{tabular}
\caption{List of additional decay channels relevant for dielectron
production in high-energy collision systems.
} \label{adddilcha}
\end{table}

We have considered all $K^0$'s as fixed 50\% - 50\% mixtures of
$K_S^0$ and $K_L^0$ states and ignored the time dependent
neutral kaon oscillations. Making a distinction between the
long and short living states is important since the $K_L^0$ has roughly
a factor of 100 larger probability to decay semileptonically than the $K_S^0$.
Electrons and positrons stemming from $K_L^0$ decays are taken into account only
if the $K_L^0$ has decayed before the first detector (2 meters from the 
primary vertex) which reduces significantly the di-electron yields from 
the $K^0$ decays.

\begin{figure}[!ht]
\includegraphics[angle=0,scale=1.0]{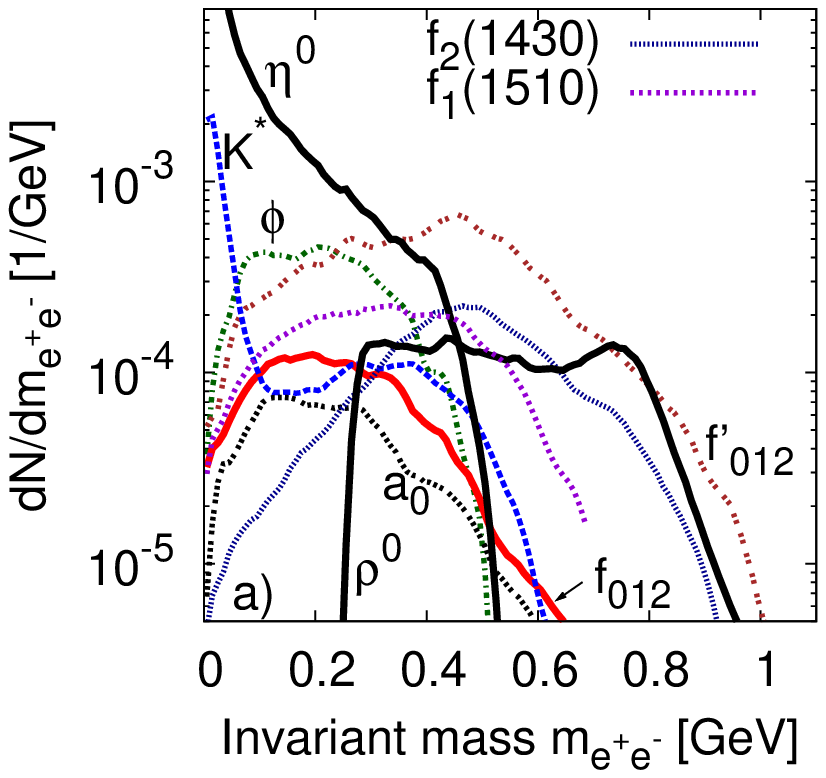}
\includegraphics[angle=0,scale=1.0]{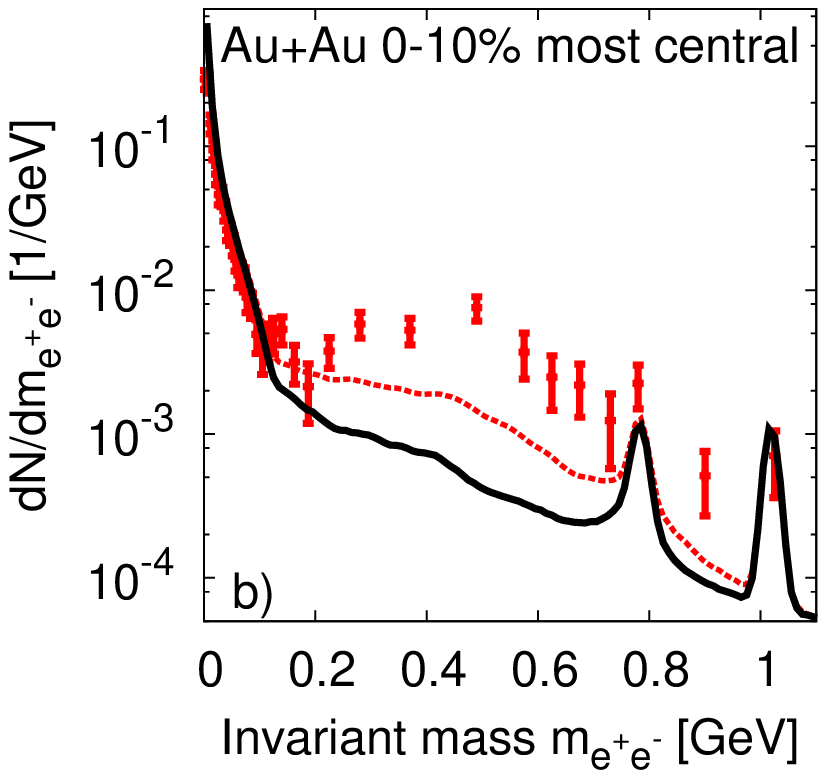}
\caption{(Color on-line) The upper limits for the invariant mass spectrum 
of pairs of electrons and
positrons in (0-10\%) most central $Au+Au$ collisions arising from
various correlated sources (top panel) in comparison with the
PHENIX data and the standard cocktail calculation (bottom panel).
\label{correlated}}
\end{figure}

The contributions from the additional correlated channels in (0-10\%) 
most central $Au+Au$ collisions are shown in top panel
of Figure~\ref{correlated}. We have added some of the channels in
Figure~\ref{correlated} for clarity. The $K^{*}$ denotes the sum of 
$K^{*\pm}$, $K^{*0}$ and $\bar{K^{*0}}$ while $f_{012}$ is the sum of 
$f_0$, $f_1$ and $f_2$ and similarly $f'_{012}$ denotes the sum of the 
$f_0(1370)$, $f_1(1420)$ and $f_2(1525)$. The contribution from the 
$f_0(1500)$ is small and is not shown. For comparison, also the hadronic 
cocktail contribution from $\eta$ and $\rho^0$ mesons are shown by the 
thick solid lines. One can see that the correlated channels might indeed give 
a sizeable contribution to the dielectron invariant mass spectrum
in the low invariant mass region and might even over-shine the standard hadronic
cocktail emission precisely in the invariant mass region where the large
excess was measured by PHENIX in central nucleus-nucleus collisions.

In the bottom panel of Figure~\ref{correlated} the
standard hadronic cocktail result (solid) and the cocktail + additional
correlated emission (dashed) are compared with the PHENIX data in the
most central  bin. Our results show that the correlated
background from the exotic mesonic states may result in a large enhancement
of the dielectron yields in the LMR, although re-scattering effects could
significantly alter the results at least for some of the correlated channels.
The sizeable contribution - relative to proton-proton reactions -
is due to a factor $\gamma_S^2$ which increases roughly by a
factor of three from peripheral to central nucleus-nucleus
collisions. Note, however, that we have addressed upper
limits, only, and that a decorrelation by interactions in the hot
medium is expected. Nevertheless, even our upper limit is clearly
below the PHENIX signal for central nucleus-nucleus collisions and
we may conclude that the additional channels considered here
should not be responsible for the dilepton excess seen
experimentally.

\subsubsection{Intermediate mass region}

We have estimated the charmed hadron yields in
heavy-ion collisions by multiplying the corresponding yields in
proton-proton collisions with the number of binary collisions for
the contributions from charm mesons. This procedure should hold
for open charm hadrons but actually we know that the production of
charmonium states suffer suppression in the hot and dense
environment of partonic and hadronic nature. This suppression is
usually expressed by the ratio $R_{AA}=\frac{\d N_x^{pp}}{\Nbin \d
N_x^{AA}}$ which parametrizes the deviation from the simple
$\Nbin$ scaling. PHENIX has measured this quantity for \jpsi~'s both in
$Au+Au$~\cite{Adare:2006ns} as well as in $Cu+Cu$~\cite{Adare:2008sh}
collisions.
So far there is no fully
convincing  model calculation that explains the observed $R_{AA}$
on a satisfactory level and thus we have taken the simplest
approach and assumed that $R_{AA}$ is a linear
function\footnote{
$R_{AA}^{AuAu}=0.75 - 0.00137\NP$ \\
$R_{AA}^{CuCu}=1.0 - 0.00516\NP$}
of centrality.
This approximation seems to hold sufficiently far away
from the ends of the whole centrality range. For alternative
curves for the $R_{AA}^{{\mathrm J}/\psi}$ with centrality we refer
the reader to the review \cite{Linnyk:2008hp}.

The different model results in Figs. \ref{fig7} and \ref{fig6}
take into account the charmonium suppression. It is clear from
Fig. \ref{fig6} that it is necessary and sufficient to include the
charmonium suppression effects in order to describe the \jpsi~peak
correctly at all centralities. Since the $R_{AA}$ ratios of the
excited charmonium states are not yet measured, we have not
implemented any correction for the $\psi'$.

Let us now turn our attention to the slope between the $\phi$ and
\jpsi~peaks. From the discussion above it is clear that both the
very low invariant mass ($\pi^0$ Dalitz decay) as well as high
invariant mass ($J/\Psi$) regions are fairly well described by the
model. This is a good starting point to address the physics of the
IMR in between.

In $p+p$ collisions there is no medium that would distort the
correlation among the emitted open charm mesons and indeed, we
have seen that the $p+p$ data can be best described by assuming
strong correlations among the \D~mesons. In central heavy-ion
collisions on the other hand, there are several hundreds of hadrons
emitted in each event and one would expect that the produced
charmed hadrons interact with the surrounding medium thus
destroying the initial correlations. We have studied the
reinteractions of the charm mesons quantitatively within the HSD
transport approach (cf. Ref. \cite{Linnyk:2008hp}) and have
calculated the probability that neither of the two \D~mesons
interacts with the surrounding medium. In this and only in this
case the angular correlations among the emitted electrons and
positrons would be preserved and remain similar to the $p+p$ case.

\begin{figure}[!ht]
\includegraphics[angle=0,scale=1.0]{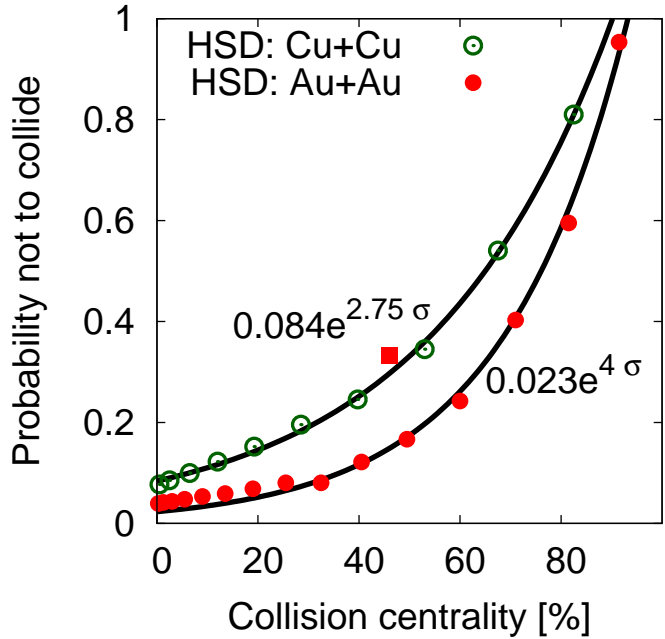}
\caption{(Color on-line) Probability that neither of the \D~and
$\bar{D}$ mesons collide with the surrounding medium in $Cu+Cu$
(open circles) and $Au+Au$ (closed circles) collisions at
\sqrtsnn=200 GeV as evaluated with the HSD transport approach. The
filled square denotes the value for minimum bias $Au+Au$
collisions. The lines are parametrization that we have
implemented in our calculations. \label{survprob}}
\end{figure}

Our results from the HSD calculations are shown in Fig.
\ref{survprob}. The probability that the angular correlations
remain the same as in $p+p$ collisions is calculated as a function
of collision centrality in $Cu+Cu$ (open circles) and in $Au+Au$
(filled circles) collisions at \sqrtsnn=200 GeV. For practical
purposes, we have parametrized these probabilities by the
functions shown in the Figure. In the case of minimum bias $Au+Au$
collisions, a proper weighted average over the whole centrality
range (denoted by the filled square in Fig. \ref{survprob}) is
used in our calculation instead  of the explicit parametrization.

We can now study the invariant mass spectrum of the dilepton
continuum in heavy-ion collisions in a more realistic scenario in
which the slope of the IMR arises as a superposition of correlated
and un-correlated open charm decays. All the three cases are shown
for the minimum bias $Au+Au$ collisions in the lower panel of Fig.
\ref{fig7}. The result retaining the correlations like in $p+p$
collisions is shown by the double dotted line in Fig. \ref{fig7}b,
while the random correlation case is represented by a dashed line.
The solid line in between shows the invariant mass spectrum
assuming that in 33\% of the cases the dilepton pair stemming from
the open charm decays retains the initial correlations while in
67\% of the cases at least one of the \D~mesons has scattered and
thus the correlation is destroyed. One can see that the most
realistic case naturally interpolates between the two extreme
cases and seems to describe the experimental data best (within the
error bars).

Some further information on the IMR is gained by having an
explicit look at the centrality dependence of the dilepton yield
for  $Au+Au$ and $Cu+Cu$ collisions at the top RHIC energy. The
solid lines in Figure \ref{fig9} are evaluated in this mixed
scenario - based on the HSD rescattering calculations - and one
can see that this approach practically underestimates the
experimental spectra in the IMR for both systems.

Presently, we
may only speculate that there seem to be further channels of
possibly partonic nature in the IMR as suggested by several groups 
independently. Within thermal models this excess might be addressed as 
thermal dilepton radiation from the QGP~\cite{Shuryak:1978ij} 
while also hadronic $\rho + \rho$ or $\pi + a_1$ scattering might
contribute as suggested by van Hees and 
Rapp~\cite{vanHees:2007th,vanHees:2006ng}. The studies by one of the authors
on explicit partonic reaction channels in 
Refs. \cite{olena10,Linnyk:2010ub} allow for an 
implementation in the PHSD transport approach \cite{PHSD} which hopefully 
might clarify this issue in the near future.

In addition to the change in the angular correlations among the open
charm mesons, also the magnitude of the (transverse) momentum of the
\D~mesons can change due to interactions in the fireball. The PHENIX
collaboration has measured the $R_{AA}$ of single electrons coming from
heavy quark decays~\cite{Adare:2010de} in $Au+Au$ collisions and found that
this $R_{AA}$ is compatible with unity up to $p_T$ of 2 GeV at all
centralities. Thus, heavy-quark energy loss can (significantly) modify the
di-electron yields in the IMR coming from \D~meson decays only at transverse
momenta larger than 2 GeV.

\begin{figure}[!ht]
\includegraphics[angle=0,scale=1.0]{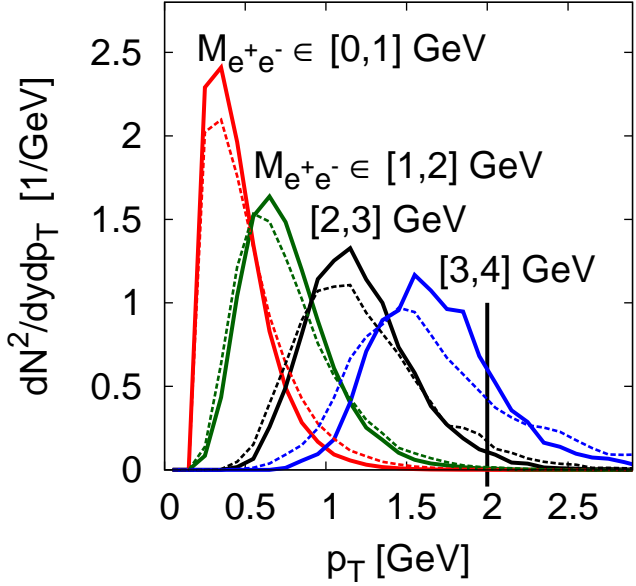}
\caption{(Color on-line) Transverse momentum distribution of electrons
and positrons (within the PHENIX acceptance, normalized to unity) coming from
semi-leptonic decays of
\D~mesons in four different invariant mass windows $M_{e^+e^-}$. The solid
lines indicate the spectra in the fully correlated case while the dashed
lines arise in the random correlation picture. Electrons and positrons
with $p_T$ larger than 2 GeV (right from the solid vertical line) exhibit
energy loss while propagating through the hot and dense
medium~\cite{Adare:2010de}.
\label{electronTspectra}}
\end{figure}

\begin{figure}[!ht]
\includegraphics[angle=0,scale=0.98]{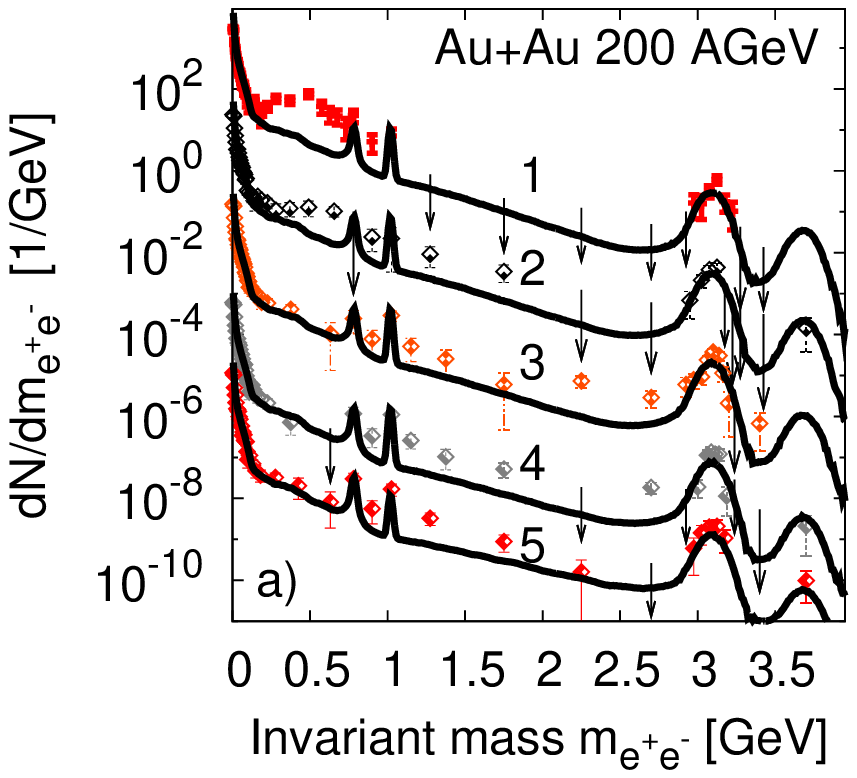}
\includegraphics[angle=0,scale=0.98]{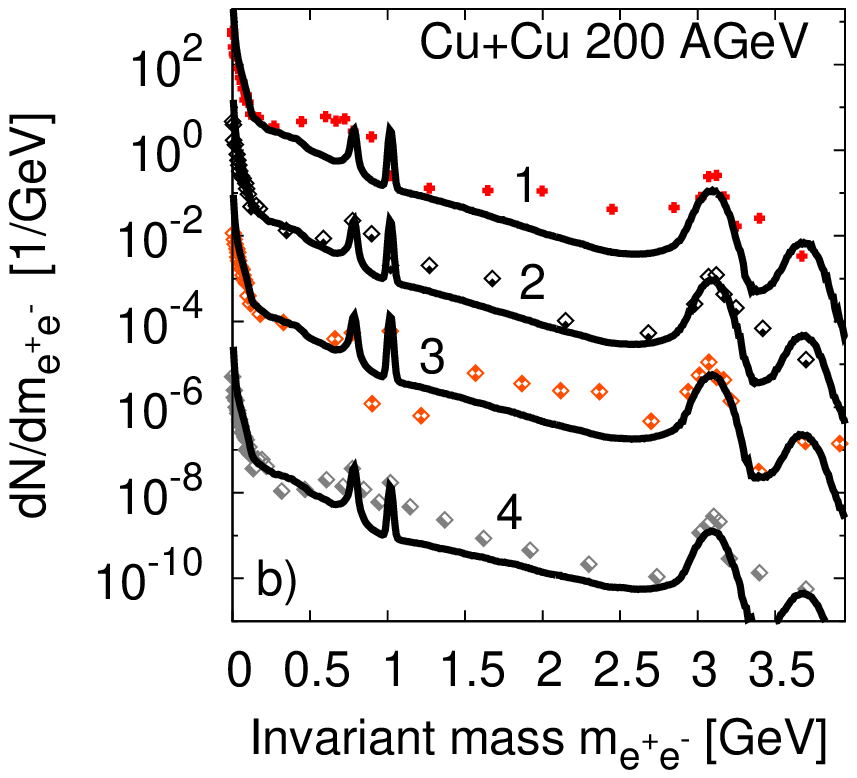}
\caption{(Color on-line) Invariant mass spectrum of pairs of
electrons and positrons in $Au+Au$ (top)~\cite{Adare:2009qk} and
$Cu+Cu$ (bottom)~\cite{Campbell} collisions at \sqrtsnn=200 GeV in
different centrality classes compared with model calculations. The
centrality bins are labeled from central to peripheral as 1
(0-10\%), 2 (10-20\%), 3 (20-40\%). Centrality bin 4 consist of
(40-94\%) and (40-60\%) most central events in $Cu+Cu$ and $Au+Au$
collisions, respectively, while the centrality bin 5 includes
(60-92\%) most central collisions. Both the data and model
calculational results are scaled with factors of $10^4$ (0-10\%),
$10^2$ (10-20\%), $1$ (20-40\%), $10^{-2}$ (40-60\%) as well as
$10^{-3}$ (60-92\%) and (40-94\%) for clarity. The solid lines
indicate the model results scaled from $p+p$ collisions such that
the charmed hadron yields are scaled with $N_\mathrm{bin}$ while
the non-charmed hadron yields are scaled with $\NP$ ($Cu+Cu$) and 
with the fitted volume ($Au+Au$) while
correlations among the open charm hadrons are evaluated in the
mixed procedure as described in the text. The starting point of the downward
pointing arrows denote the data points which are defined as upper limits only.
\label{fig9}}
\end{figure}

In order to estimate how this affects our results, we have
calculated the single electron + positron transverse momentum spectrum
within the PHENIX acceptance and divided the results in four classes according
to the corresponding invariant mass of the di-electron pair $M_{e^+e^-}$.
The $p_T$ spectra are shown in mass windows of $M_{e^+e^-} \in$
[0,1] ; [1,2] ; [2,3] and [3,4] GeV in Figure~\ref{electronTspectra}. The
solid lines indicate the results in the fully correlated case while the
dashed lines are evaluated in the random correlation picture. All curves
are normalized to unity. One can see
that a significant fraction of electrons and positrons with $p_T>$2
GeV contributes to the invariant mass spectrum at invariant masses larger
than 3 GeV, only. Our calculations concentrate on the region
$M_{e^+e^-} \in $ [0,$M_{J/\psi}$] and thus we can conclude that our results
are only little affected by the heavy-quark energy-loss effects in the medium.
In particular, the invariant mass region $M_{e^+e^-} \in$ [1,2] GeV - in which
our calculations under-estimate the measurements - appear little affected by
the heavy-quark energy loss.

\section{Discussion}

As we have seen in the previous Section a scaling of the
dielectron yields from $p+p$ to heavy-ion collisions can lead to a
surprisingly good description of the data in the peripheral
collision systems but fails for the more central collisions. The
question we wish to address here is that if we can, nevertheless,
understand the observed excess in terms of hadronic degrees of
freedom or if additional partonic productions channels have to be
incorporated.

According to the statistical hadronization model fits to $p+p$ and
$Au+Au$ collisions at RHIC, the intensive thermal characteristics
of these systems seem very similar at mid-rapidity. All of the
light mesons decaying into dileptons are completely neutral and so
the dilepton production rate does not actually depend on the
chemical potentials at all and thus, besides the $\gamma_S$
parameter discussed before, the temperature is the only
intensive parameter left in the model that could lead to a
non-trivial scaling behavior seen in the data. However, one of the
lessons we have learned from the SHM fits to RHIC
data is that the temperature is the same in heavy-ion collisions
at all centralities and this temperature coincides with the one
extracted from the $p+p$ collision data. We have, nevertheless,
checked that a moderate increase ($T=170\rightarrow 190$ MeV) in
temperature can not explain the observed excess in the LMR. One
should notice that due to the momentum cut $p_T>0.2$ GeV the
increase in temperature affects  more prominently the
dilepton yields from the vector mesons than the emission from
$\pi^0$ and thus a change in temperature affects different regions
in the invariant mass spectrum with different strength. One would
nevertheless need unrealistically large temperatures of $T>200$
MeV if one attempts to assign the dilepton excess (seen by the PHENIX 
Collaboration) to an increase of the fireball temperature. We rule out such a
possibility.

Independent previous model
calculations~\cite{lena2009,Dusling,vanHees:2007th}
have been compared with
the PHENIX data in the original publication~\cite{Adare:2009qk} and we
refer the reader to Figs. 41 and 42 of Ref.~\cite{Adare:2009qk} for
details. Our results agree qualitatively and also quantitatively
with the previous model calculations in that the proton-proton
collisions are well described in the whole invariant mass range
while none of the analyses can explain the excess in the central
heavy-ion collisions in the low invariant mass region. In
extension of the previous studies we have, furthermore,
investigated the possibility that the observed excess might stem
from further semileptonic correlated kaon decays which are
enhanced in central nucleus-nucleus collisions relative to
proton-proton reactions by roughly a factor of three. Our upper
limits for the dominant channels considered here clearly show that
also these additional 'background sources' are not responsible for
the large excess seen by the PHENIX Collaboration in central
heavy-ion reactions.

The charmed sector or the IMR (if considered) has been treated
essentially in a similar fashion by the PHENIX Collaboration as in this work and no solid
conclusions have been possible, so far. In order to go beyond the
previous attempts we have calculated the rescattering
probabilities of charm mesons dynamically (within HSD) which
allows to estimate the amount of uncorrelated electron + positron
pairs from \D~meson decays as a function of the centrality of the
reaction. Our final results for $Au+Au$ and $Cu+Cu$ suggest that
we clearly underestimate the preliminary yield from PHENIX which
might point towards partonic sources - as suggested in Ref. 
\cite{Linnyk:2010ub} - in the intermediate mass regime.

So far we have considered only events with exactly one charmed
quark pair. Processes leading to un-even amounts of charmed quarks
are possible but they are more rare than the case we have studied
and the corrections are probably not very large. Events with, e.g.
3 \D~mesons tend to populate the low invariant mass region, in
which the open charm contribution is insignificant, because in
that case only one of the possible two ''dilepton''-pairs is
(strongly) correlated. To finally clear up the situation we are going 
to carry out
non-perturbative calculations on correlated charm dynamics within
the PHSD transport approach \cite{PHSD} that also includes the
dynamics of charm quarks in the partonic phase.

\section*{Acknowledgments}
The authors like to thank A. Toia for stimulating discussions.
Furthermore, E.L.B. and O.L. are grateful for financial support
from the 'HIC for FAIR' center of the 'LOEWE' program and J.M. for
support from DFG.

\end{document}